\input harvmac
\input epsf
\noblackbox
\def\IP{{\bf P}}\def\IZ{{\bf Z}}\def\IR{{\bf R}}
\def\br{\hfill\break}\def\ni{\noindent}\def\ov#1#2{{#1 \over #2}}
\def\cx#1{{\cal #1}}\def\al{\alpha}\def\IP{{\bf P}}
\def\br{\hfill\break}\def\ni{\noindent}\def\ov#1#2{{{#1}\over{#2}}}
\def\ep{\epsilon}\def\al{\alpha}\def\be{\beta}
\def\zg{z^{(g)}}\def\zm{z^{(m)}}\def\zc{z^{(c)}}

\def\lm#1#2{{l_{#2}^{(#1)}}}
\def\subsubsec#1{\ni{\it #1}\br\ni}
\def\IF{{\bf F}}\def\ICs{{\bf C}^\star}
\def\xs{X^\star}\def\IR{\bf R}
\def\la{\lambda}\def\IC{{\bf C}}\def\vt{\tilde{v}}
\def\at{\tilde{a}}\def\xt{\tilde{x}}
\def\WP{{\bf WP}}\def\yc{y^{(c)}}\def\yg{y^{(g)}}\def\zp{z^\prime}
\def\mprod#1#2{{\langle {#1} , {#2} \rangle}}
\def\ns#1{{\nu_{#1}^\star}}\def\dd{\bigtriangledown}
\def\nst#1{{\tilde{\nu}_{#1}^\star}}
\def\us#1{\underline{#1}}
\def\nup#1({Nucl.\ Phys.\ $\us {B#1}$\ (}
\def\plt#1({Phys.\ Lett.\ $\us  {B#1}$\ (}
\def\cmp#1({Comm.\ Math.\ Phys.\ $\us  {#1}$\ (}
\def\prp#1({Phys.\ Rep.\ $\us  {#1}$\ (}
\def\prl#1({Phys.\ Rev.\ Lett.\ $\us  {#1}$\ (}
\def\prv#1({Phys.\ Rev.\ $\us  {#1}$\ (}
\def\mpl#1({Mod.\ Phys.\ Let.\ $\us  {A#1}$\ (}
\def\ijmp#1({Int.\ J.\ Mod.\ Phys.\ $\us{A#1}$\ (}
\def\tit#1|{{\it #1},\ }
\lref\gross{P. Aspinwall and M. Gross, \plt 387 (1996) 735}
\lref\ber{M. Bershadsky et. al., \nup 481 (1996) 215}
\lref\kv{S. Katz and C. Vafa, {\it Matter from geometry},
hep-th/9606086}
\lref\fmw{R. Friedman, J. Morgan and E. Witten, {\it Vector
bundles and F theory}, hep-th/9701162}
\lref\loo{E. Looijenga, Invent.  Math.  {\bf 38} (1977) 17;
Invent. Math. {\bf 61} (1980) 1.}
\lref\morI{D. R. Morrison and M. R. Plesser, \nup 440 (1995) 279}

\Title{\vbox{
\hbox{HUTP-97/A025}
\hbox{OSU-M-97-5}
\hbox{IASSNS-HEP-97/65}
\hbox{\tt hep-th/9706110}
}}
{}
\vskip-3cm
\centerline{\titlefont Mirror symmetry}
\vskip0.3cm
\centerline{{\titlefont and}}
\vskip0.4cm
\centerline{\titlefont {Exact Solution of 4D $N=2$ Gauge Theories\ --\ I}}
\bigskip\bigskip
\centerline{Sheldon Katz$^{a,b}$, Peter Mayr$^{c}$
and Cumrun Vafa$^{d}$}
\vskip .2in
\centerline{\it $^{a}$
Oklahoma State University, Stillwater, OK 74078, USA}
\vskip.05in \centerline{\it $^{b}$Institut Mittag-Leffler,
S-182 62 Djursholm, Sweden}
\vskip.05in \centerline{\it $^{c}$School of Natural Sciences, Institute for
Advanced Study, Princeton, NJ 08540, USA}
\vskip.05in \centerline{\it $^{d}$Lyman Laboratory of Physics, Harvard
University,
Cambridge, MA 02138,USA}
\vskip .3in
\noindent
Using geometric engineering in the context
of type II strings, we obtain exact solutions for the moduli space
of the Coulomb branch of all $N=2$ gauge theories
in four dimensions involving products of $SU$ gauge
groups with arbitrary number of bi-fundamental matter
for chosen pairs, as well as
an arbitrary number of fundamental matter for each factor.
Asymptotic freedom restricts the possibilities to $SU$ groups
with bi-fundamental matter chosen according to ADE or affine ADE
Dynkin diagrams.  Many of the results can be derived
in an elementary way using the self-mirror property of $K3$.
We find that in certain
cases the solution of the Coulomb branch for $N=2$ gauge theories is
given in terms of a three dimensional complex manifold
rather than a Riemann surface. We also study new stringy strong coupling
fixed points arising from the compactification of higher dimensional
theories with tensionless strings and consider applications to three
dimensional $N=4$ theories.
\Date{June 1997}


\newsec{Introduction}
Many non-trivial facts involving exact results
for supersymmetric field theories have found their natural
explanation in the context of string theories.  In particular
many exact results about field theories can be obtained from
a realization of them by considering special limits of string
compactifications and the use of classical string symmetries,
such as T-duality or mirror symmetry.

In this paper we concentrate on the case of $N=2$ quantum field theories
in $d=4$
and obtain exact results for the Coulomb branch of such theories by using
classical symmetries and in particular mirror
symmetry of type II string compactifications
on Calabi-Yau threefolds.
Exact $N=2$ results from string theories
were first obtained in
\ref\kacv{S. Kachru and C. Vafa, \nup 450 (1995) 69}\
by conjecturing an exact duality between heterotic strings
on $K3\times T^2$ and certain Calabi-Yau compactifications of type II
strings.  It was pointed out in
\ref\klm{A. Klemm, W. Lerche and P. Mayr, \plt 357 (1995) 313}\
that the relevant Calabi-Yau's which arise in these dualities are
of the form of a $K3$ fibration over a 2-sphere.  It was argued
in
\ref\vw{C. Vafa and E. Witten, {
\it Dual string pairs with N=1 and N=2 supersymmetry in
four-dimensions}, hep-th/9507050}\
that this has a natural interpretation
based on fibering the duality of heterotic string on $K3$ with type
IIA strings on $T^4$.  In fact it was shown in
\ref\las{J. Louis and P. Aspinwall, \plt 369 (1996) 233}\
that $K3$ fibration of Calabi-Yau threefold is
a necessary condition for any $N=2$ type II/heterotic duality.
After a careful study of the field theory content of the string duality
conjectured in \kacv\ it was found in
\ref\kklmv{S. Kachru et. al., \nup 459 (1996) 537}\
that the relevant field theory part of the moduli comes from an
ADE type singularity of the $K3$ fibered over the sphere ${\bf P^1}$.
This resulted in a derivation of many of the exact results known
for $N=2$ field theories in four dimensions
\ref\ne2{N. Seiberg and E. Witten, \nup 426 (1994) 19, erratum: ibid
$\underline{B 430}$ (1994) 396;
\nup 431 (1994) 484.}\ref\orsw{A. Klemm, W. Lerche, S. Theisen, and S.
Yankielowicz,
\plt 344 (1995)\semi
P. Argyres and A. Faraggi, \prl 73 (1995) 3931 \semi
A. Hanany and Y. Oz, \nup 452 (1995) 283\semi
P. Argyres, M. Plesser, and A. Shapere, \prl 75 (1995) 1699\semi
U. Danielsson and B. Sundborg, \plt 358 (1995) 273\semi
A. Brandhuber and K. Landsteiner, \plt 358 (1995) 73\semi
A. Hanany, \nup 466 (1996) 85\semi
P. C. Argyres and A. D. Shapere, \nup 461 (1996) 437\semi
W. Lerche and N. Warner, {\it Exceptional SW Geometry from
ALE Fibrations}, hep-th/9608183\semi
K. Landsteiner, J. M. Pierre, S. B.
Giddings, \prv D55 (1997) 2367\semi
E. Martinec and N. Warner, \nup 459 (1996) 97\semi
E. D'Hoker, I.M. Krichever and D.H. Phong, \nup 489 (1997) 211}
\lref\witex{E. Witten,
{\it Solutions of four-dimensional field theories via M theory},
hep-th/9703166}
\lref\witexf{
K. Landsteiner, E. Lopez and David A. Lowe
{\it N=2 supersymmetric gauge theories, branes and
                  orientifolds}, hep-th/9705199\semi
A. Brandhuber, J. Sonnenschein, S. Theisen and S.
                  Yankielowicz, 
{\it M theory and Seiberg-Witten curves: Orthogonal and
                  symplectic groups}, hep-th/9705232}.
Moreover it was shown in
\ref\klmvw{A. Klemm, W. Lerche, P. Mayr, C. Vafa and N. Warner,
\nup 477 (1996) 746}\
(for a review of various aspects of this construction
see \ref\lercr{W. Lerche,{\it
Introduction to Seiberg-Witten theory and its stringy
                  origin},hep-th/9611190\semi
A. Klemm, 
{\it On the geometry behind N=2 supersymmetric effective actions
                  in four-dimensions}, hep-th/9705131}) how
to use this string description and compute directly the BPS
spectrum of $N=2$ strings in terms of geodesics on the Seiberg-Witten
curve.  This study of BPS states has now been
extended to many cases
\ref\warn{A. Brandhuber and S. Stieberger, \nup 488 (1997)
177\semi
J. Schulze and N. P. Warner,{
\it  BPS geodesics in N=2 supersymmetric Yang-Mills theory},
hep-th/9702012\semi
J. M. Rabin, {
\it Geodesics and BPS states in N=2 supersymmetric QCD},hep-th/9703145
}.
It was also shown in \klmvw\ using the T-duality
between NS 5-branes and ADE singularities
\ref\oov{H. Ooguri and C. Vafa, \nup 463 (1996) 55}
that one can view the four dimensional theory as coming from the
study of the type IIA 5-brane whose worldvolume is $R^4\times \Sigma$
where $\Sigma$ is a non-compact version of the Seiberg-Witten curve.

The analysis in \klmvw\ hinted that the basic
idea of embedding $N=2$ field theories in type II
compactifications on Calabi-Yau threefolds, can be done
{\it without appealing to any non-trivial string duality}.  Namely
just using the fact that in type IIA D2 branes wrapping over
vanishing spheres of ADE singularity give rise to gauge symmetries
in six dimensions, and fibering that over a sphere giving $N=2$
theories in four dimensions gives by itself a complete picture.
This idea was further developed in
\ref\kkv{S. Katz, A. Klemm and C. Vafa, {
\it Geometric engineering of quantum field theories},
hep-th/9609239}\
where
it was called {\it geometric engineering} of quantum field theories.
In that context one starts with type IIA strings on Calabi-Yau threefolds,
performs a local mirror transformation, and ends up with a local type IIB
model.
Moreover the moduli of the complex structure mirror type IIB is not corrected
quantum mechanically and thus gives the exact quantum answer already at
the type IIB string {\it and} the worldsheet tree level.
The main aim of this paper is to continue the idea of geometric engineering
into a much wider class of $N=2$ theories.  Since there are many
possibilities to consider, we have decided to devote two papers to this
subject.
In the first paper (the present one) we discuss geometric
engineering of $N=2$ gauge theories involving products of $SU$ groups
with bi-fundamental matter, as well as extra fundamental matter.
We provide exact solution for the Coulomb branch moduli for
{\it all} such cases.
The various cases will
depend on the configuration of matter we consider
which allows us to attach a ``quiver diagram'' associated to our theory.
In particular we consider a diagram where for each $SU$ gauge
group we consider a node, and for each pair of groups with bi-fundamental
matter, we connect the corresponding nodes with a line.  It turns
out, as we will show, that asymptotic freedom in four dimensions implies
that the corresponding diagram corresponds to ADE Dynkin diagram
or affine ADE Dynkin diagram.  In the case we get ADE Dynkin diagram
we can add extra fundamental matter to make the theory superconformal.
In the case of affine ADE the condition of having asymptotic
freedom will imply that the rank of each $SU$ gauge group
is correlated with the Dynkin number on the corresponding node,
and that automatically leads to a superconformal theory (without
any extra fundamental matter).  As we will show, the S-duality
group in all these cases corresponds to the fundamental
group of the moduli of flat ADE gauge
fields on elliptic curve or the degenerate elliptic curve,
depending on whether we are dealing with the affine ADE Dynkin
diagram or the ordinary ADE Dynkin diagram.  It is quite surprising
from the field theory perspective how the ADE gauge fields appear
in this story.  We will explain its stringy origin.

We construct the type IIA geometry and its type IIB mirror for all
 such theories.  The
complex geometry of the mirror, which gives the exact solution
for the moduli of Coulomb branch, is generally given by a non-compact piece of
 a Calabi-Yau 3-fold.  Sometimes, {\it but not always}, we find
that the data can be reduced to a Riemann surface $\Sigma$,
which as noted in \klmvw\ becomes equivalent by a T-duality \oov\
to the fivebrane
of type IIA (or M-theory) on $R^4\times \Sigma$.

In a subsequent paper we generalize these constructions
to include more general gauge groups and matter content.

The organization of this paper is as follows:
In section 2 we introduce the basic idea of geometric engineering
of $N=2$ theories in type IIA strings and the strategy
of getting exact results by application of mirror symmetry.
We also show how asymptotic freedom restricts the bi-fundamental
matter structure to be in the form of ADE or affine ADE Dynkin
diagrams.
In section 3 we discuss some relevant intuitive
aspects of mirror symmetry, and in particular the fact that
ALE space is self-mirror,
to help us develop an intuition about the results
that are to follow.  Moreover, in that section we give a simple, but heuristic
derivation of some of the results (which are derived later in the
paper using more sophisticated and rigorous
toric methods).  These include the case of
$SU$ gauge groups arranged along a linear chain
with bi-fundamental matter between nearest neighbors (the A case),
and the case of $SU$ with bi-fundamental
matter according to links of
affine $E$ Dynkin diagram.
 In section 4 we discuss aspects of toric
geometry and its relation to mirror symmetry.  The discussion
in this section is aimed at putting the intuitive arguments
in section 3 in the powerful setup of toric geometry. We aim
to provide an essentially self-contained introduction to toric
geometry and its relation to mirror
symmetry, emphasizing aspects which we will use
in this paper.
In section 5 we revisit the case of
$SU$ gauge groups arranged along the linear chain (the
A case)
with bi-fundamental matter and rederive the results
of section 3 more rigorously.  This also allows us to give
more information about the solution including the relevant
meromorphic 1-form on the Riemann surface.
In section 6 we show how arbitrary matter in the fundamental
representation
can be added for each gauge factor.  This construction
will be applicable to all ADE cases.  In section 6 we mainly concentrate
on the A case with extra matter.
In section 7 we discuss
a quiver configuration corresponding to a trivalent vertex.
This in particular allows us to rederive the results concerning
the affine $E$ cases discussed in section 3 more rigorously.
We also show how the theories
based on ordinary Dynkin diagrams of $D$ and $E$ are realized
in this construction.
In section 8 we give a uniform treatment for quivers
based on affine ADE Dynkin diagrams, by studying the mirror
of elliptic ADE singularities in 2-complex dimensions
(this gives the first derivation for the affine A and D cases and a third
derivation for the affine E case as the quiver).  We also
discuss the S-duality group for these affine ADE
cases as well as the case based
on ordinary ADE Dynkin diagrams.  In particular we find that the S-duality
groups for all these cases have a rather simple description in terms
of the duality group for ADE flat connections on a two dimensional
torus and its degeneration.
In section 9  we discuss some realization of certain
stringy strong coupling
fixed points, which can be solved using our method.
 In particular we show that in some cases we get new critical
$N=2$ theories (which are related to toroidal compactifications
of some tensionless non-critical string theories
found in six dimensions).  In section 10
we apply our results to obtain some new results
for $N=4$ theories in $d=3$.  In particular
we construct the dual of $k$ small instantons of $E_8$ theory
compactified to three dimensions, extending the result
for $k=1$ obtained in
\ref\hov{K. Hori, H. Ooguri and C. Vafa,{
\it Non-Abelian conifold transitions and N=4 dualities in three-dimensions},
hep-th/9705220}\
and verifying the conjecture in
\ref\inse{K. Intriligator and N. Seiberg, \plt 387 (1996) 513}.

\newsec{Basic Setup}
Our starting point is a local model for type IIA compactification
on a non-compact Calabi-Yau threefold.  We first need to review
some facts about properties of type IIA strings on ALE spaces
(some of which arise as local singularities of $K3$ manifold).
Consider an ALE space with an ADE type singularity.  For simplicity
let us consider the $A_1$ case.  In this case we have a singularity
of the form
\eqn\aone{xy+z^2=0.}
This singularity can be resolved by `blowing up'.  Concretely
what this means is that we consider a new variable
$${\tilde x}=x/z$$
which implies that if we substitute it into the above equation it
is of the form
$${\tilde x}y+z=0$$
which is not singular any more. This resolution
has been at the price of doing a singular
change of variables. Mathematically what we have done
is to replace $x=z=0$ which was the point singularity of the
original space by a whole sphere parametrized by $\tilde x$, and having
done that we have avoided the singularity.  We are only describing
the complex structure of the curve, but if we wish to put metrics
to make this resolution continuously match with the singular
manifold we started with, we have to make the sphere denoted by
${\tilde x}$ have zero volume at the beginning
and then increase it continuously to a finite
value.  In the context of type IIA string propagating
on this background, D2 branes wrapped around the $\tilde x$ sphere
will give a vector particle whose mass is proportional
to the volume of the blown up ${\bf P}^1$ (2-sphere).  Actually we can
have two different orientations for the wrapping of the D2 brane
and so we obtain two states, which we will denote by $W^{\pm}$.
The states $W^{\pm}$ are charged under the $U(1)$ gauge field
corresponding to decomposition of the type IIA 3-form
in terms of the harmonic form on the ${\bf P}^1$.  Let us call
this vector field by $Z$.  In the limit where the ${\bf P}^1$ shrinks
we get three massless vector fields $W^{\pm},Z$ which form
an $SU(2)$ adjoint.  The story is similar for the general ADE singularities
where we obtain an enhanced ADE gauge symmetry in the limit where
all the 2-cycles shrink (for a recent
review of ADE blowups in the physics literature see
\hov ). We thus obtain an $N=2$ ADE gauge
symmetry in $d=6$.

If we compactify on a $T^2$ down to $d=4$ we obtain
an $N=4$ system. Note that the
extra scalars we get in the $N=4$ system can be identified
with the expectation values of Wilson lines on
the $T^2$.
 We are however interested
in obtaining an $N=2$ system in $d=4$. In order to kill
the extra scalars we need the intermediate
two space to have no cycles, which means
that we need a 2-sphere.  Mathematically what
this means is that we have a three complex dimensional fibered
space with a two sphere as the base and the ALE space
as the fiber.  The structure of the fibration
is such that the whole three dimensional non-compact
space can be viewed as a non-compact Calabi-Yau threefold.
We have thus {\it engineered}
$N=2$ pure Yang-Mills
theory in $d=4$.  Note that the volume of the base ${\bf P}^1$
is related to the coupling constant of Yang-Mills in $d=4$
by the usual volume factor, namely
\eqn\cov{V_{base}={1\over g^2}}
The Coulomb parameters of the $N=2$ system in $d=4$
get mapped to the sizes of the blown up ${\bf P}^1$'s
in the fiber.  Sometimes when we refer to the
fiber geometry we only concentrate on the compact
parts of it, namely the blown up spheres.

\subsec{Incorporation of Matter}
There have been a number of works which relate
how matter arises from the geometry of Calabi-Yau compactifications.
We will follow the approach in \kv\
which itself was based on earlier works
\ref\rellit{ M. Bershadsky, V. Sadov and C. Vafa, Nucl. Phys. B463 (1996)
398\semi
A. Klemm and P. Mayr,  \nup 469 (1996) 37
\semi S. Katz, D. R. Morrison and M. R.
Plesser, \nup 477 (1996) 105\semi
P. Berglund, S. Katz, A. Klemm, P. Mayr,
\nup 483 (1997) 209.}\ber .
For concreteness let us explain how we can obtain bi-fundamental
hypermultiplets of $SU(n)\times SU(m)$. Suppose we have an $A_{n-1}$
singularity
over ${\bf P}^1$ and an $A_{m-1}$ singularity over another ${\bf P}^1$.
Moreover the two ${\bf P}^1$'s meet at a point where the singularity
jumps to an $A_{n+m-1}$ singularity
\ref\mir{R. Miranda, {\it Smooth Models for Elliptic Threefolds},
in R. Friedman and D. R. Morrison, editors, ``The Birational Geometry
of Degenerations'', Birkh\"auser, 1983}.
Let $z_{1,2}$ denote the
coordinates of the two ${\bf P}^1$'s and assume that the intersection
point is at $z_1=z_2=0$. Consider the threefold which is locally given
by
$$xy=z_1^mz_2^n$$
Then for arbitrary $z_1\not= 0$ we have an $A_{n-1}$ singularity
and for arbitrary $z_2\not =0$ we have an $A_{m-1}$ singularity.
Let us change variables to $z_2=z_1+t$.  The above equation then becomes
$$xy=z_1^m(z_1+t)^n=z_1^{n+m}+...+t^n z_1^m $$
This can be reinterpreted as an $SU(n+m)$ singularity in 6 dimensions
which is broken to $SU(n)\times SU(m)\times U(1)$ by giving
space-dependent vevs to some of the scalars in the Cartan of $SU(n+m)$.
Recall that if we had just compactified $SU(n+m)$ on $T^2$
we would have gotten an $N=4$ system, which in the $N=2$ terms
contains an extra hypermultiplet in the adjoint.  Out of
the adjoint hypermultiplet of $SU(n+m)$ the above
space dependent breaking of $SU(n+m)\rightarrow SU(n)\times SU(m)\times U(1)$
gives rise to $(n,\bar{m})$ of $SU(n)\times SU(m)$ charged under the $U(1)$
and localized near $z_1=z_2=0$, as explained in \kv\ (see also the closely
related case \fmw).  The adjoint of $SU(n)\times
SU(m)$ that one would get in this picture has a mass because of the
global geometry of the base ${\bf P}^1$'s as explained before.
So in this way we have engineered $SU(n)\times SU(m)\times U(1)$
with bi-fundamental matter charged also under $U(1)$.  It may appear
that we are getting an `unwanted' $U(1)$.
This is not quite true. In fact we need to be able to
give an arbitrary mass to the bi-fundamental matter, and this
can be done by going to the Coulomb branch of $U(1)$.  Since the $U(1)$
is not asymptotically free we can ignore its infrared dynamics and just
think about its Coulomb branch as the mass parameters of the $SU(n)\times
SU(m)$ system.

\vskip 0.5cm
{\baselineskip=12pt \sl
\goodbreak\midinsert
\centerline{\epsfxsize 3truein\epsfbox{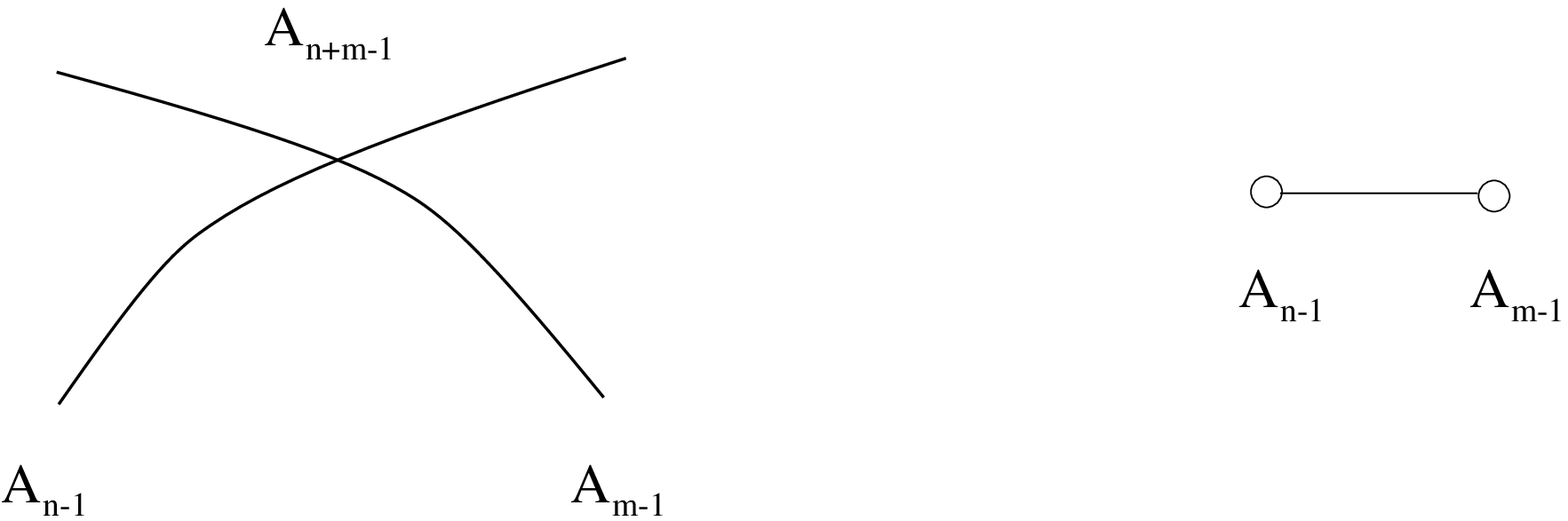}}
\leftskip 1pc\rightskip 1pc \vskip0.3cm
\noindent{\ninepoint  \baselineskip=8pt {\bf Fig. 1:}
At the intersection of two base curves carrying
$A_{n-1}$ and $A_{m-1}$ fibers, an $A_{n+m-1}$ singularity develops, with the
extra 2-sphere supporting the bi-fundamental matter. This is
denoted by a link in the quiver diagram on the right.
}\endinsert}\ni
It is now straightforward to generalize this to arbitrary product
of $SU$ groups with matter in bi-fundamentals.  The data for such
a theory can be drawn in terms of a graph, where to each gauge group
we associate a node (vertex) in the graph and for each bi-fundamental
matter between pairs of groups we draw a line connecting
the corresponding nodes.  Geometrically we engineer this
theory by associating to each node a base ${\bf P}^1$ over
which there is the corresponding $SU$ singularity, and to each pair of nodes
connected, we associate an intersection of the base ${\bf P}^1$'s, where
over the intersection point the singularity is enhanced to an $SU(n+m)$
(assuming the nodes correspond to $SU(n)$ and $SU(m)$ groups).
Note that if we are interested in addition in getting fundamental matter
for each group, this can be done by adding extra $SU$ groups with
bi-fundamental matter, roughly by gauging the flavor group, and
making the coupling constant of the extra flavor group weak,
by making the base of the corresponding ${\bf P}^1$ big
(recall \cov ).  This process we sometimes call as adding
extra nodes and `degenerating' them.

Clearly we can generalize this to more general groups in the
fiber (such as $D,E$ and non-simply laced
groups
\ref\agr{P. Aspinwall and M. Gross, \plt 387 (1996) 735}
\ref\sixau{M. Bershadsky et. al., \nup 481 (1996) 215})
and more general kind of matter (coming from the breaking
of an adjoint of a higher group) as discussed in \kv .
This more
general situation will be the subject of an upcoming paper
\ref\pape{S. Katz, P. Mayr and C. Vafa, to appear.}.

\subsec{Restrictions from Asymptotic Freedom and
ADE Dynkin Diagrams}

As noted above we are considering
the case of product of $SU(k_i)$ gauge groups with bi-fundamental
matter between some pairs and some extra fundamentals
for each group.  For interesting four dimensional field theories,
one would be interested in theories with negative $\beta$-function
for all gauge factors\foot{We will nevertheless also consider cases
with positive beta function in four dimensions, since
these four-dimensional field theories give rise to
interesting three dimensional field theories
which are asymptotically free after further
compactification on a circle \hov.}.
This turns out to put a severe restriction on the choice
of the bi-fundamental matter one chooses.  As discussed before
the structure of bi-fundamental matter gives rise to a graph
 where for each node $i$ of the graph we consider an $SU(k_i)$ gauge group
of some rank $k_i$ , and for each pair of bi-fundamentals between
the $i$-th and $j$-th group we draw a line between the $i$-th and
$j$-th node.  We will now show that the corresponding graph
is that of ADE Dynkin diagram or its affine extension.  In other
words, quite independently of what extra fundamental matter one
has, the geometry of the bi-fundamental matter is already very
restrictive\foot{We thank Noam Elkies for providing the mathematical argument
that follows.  See also \ref\vka{V. Kac, {\it Infinite dimensional
Lie algebras}, Cambridge University Press 1990}.}.

Suppose we have $r$ gauge group factors.  Let $M$ be a
symmetric $r\times r$ matrix with diagonal entries $M_{ii}=2$
and off-diagonal entries $M_{ij}=-N_{ij}$
where $N_{ij}$ is the number of bi-fundamentals
between i-th and j-th gauge groups.  Let $k$ denote the vector
which gives the rank of the gauge groups.  Then the condition
of asymptotic freedom can be succinctly stated as the requirement
$$(M k)_i=\sum_j M_{ij}k_j= 2k_i-\sum_{j\not =i} N_{ij}k_j \geq  0.$$
In other words $Mk$ is a positive semi-definite vector.
Now we need the following fact known as Perron-Frobenius theorem:

{\it If S is a symmetric matrix with positive entries, then
the eigenvector} $v$ {\it corresponding to its maximal positive
eigenvalue can be chosen to have positive entries.}

To prove this, let us normalize $v$ such that $v^t v=1$.  Then
$v$ satisfies the condition that $v^tSv$ is maximal subject to $v^tv=1$.
Consider a positive vector
$v'=|v|$.  Note that ${v^t}' v'=1$.  Since $S$ has positive entries, we
deduce
$$v'^tS v'\geq v^tS v.$$
But since $v$ maximizes $v^tS v$ then the above must be an equality and so
$v'$ should be the same
as $v$ (up to an overall sign).  In other words the eigenvector
corresponding to the maximal eigenvalue can be chosen to correspond
to a positive vector.

Now let us apply this theorem to the matrix $N$, which is a positive
symmetric matrix.  Since $M=2I-N$ where $I$ is the identity matrix,
we learn that if $v$ is a positive vector corresponding to the
maximal eigenvalue of $N$, then it is also the smallest
eigenvalue of $M$.  Let us call this eigenvalue by $\lambda$,
i.e. $Mv=\lambda v$.
Let us consider
$$v^t M k$$
Since $Mk\geq 0$, by assumption of $\beta \leq 0$, and since
$v$ is a positive vector we learn that $v^t M k$ is positive.
Thus we have
$$0\leq v^tMk=\lambda
(v^t k).$$
  Since both $v$ and $k$ are positive vectors this implies that $\lambda
\geq 0$.  Since $\lambda$, the smallest eigenvalue of $M$, is positive
this implies, as is well known, that $M$ corresponds
to an ADE  Dynkin diagram if $\lambda >0$ or corresponds
to an affine ADE  Dynkin diagram if $\lambda =0$.
Moreover, if $M$ corresponds to affine ADE Dynkin diagram,
since $v^t Mk=0$ and $v^t>0$ this implies that $Mk=0$, i.e.
we learn that $k$ is proportional to $v$ which is the only
eigenvector corresponding to zero eigenvalue for affine ADE
generalized Cartan matrix.  Note that this implies that
$k$ is a vector which is proportional to the vector
of Dynkin numbers associated to the nodes of affine Dynkin
diagram.
These interesting special cases correspond to having
ranks $k_i$ of the $i$-th $A$ factor be given by a common integer multiple
of the Dynkin labels $s_i$ of the affine ADE
$k_i= s_i n$.
In case we have ordinary ADE Dynkin diagram there
is no choice of rank vector $k$ which makes the theory superconformal
just by having bi-fundamental matter, because $Mk>0$.  In such
cases we can add $Mk$ extra fundamental matter to the corresponding
gauge group and make the theory superconformal.
These cases we will consider in more detail later.
The group $\hat{G}$ associated to the base
geometry will turn out to be of physical relevance in many respects,
such as determining the S-duality group of the conformal
four-dimensional theory, as we will discuss later in the paper.

\subsec{Strategy in Extracting Exact Results:  Mirror Symmetry}

We have described how we can geometrically engineer
$N=2$ quantum field theories in four dimensions,
in particular for $SU$ groups with bi-fundamental matter,
in the context of type IIA strings.  We are interested in using
this geometry to learn about gauge dynamics. In general
we have Higgs and Coulomb branches for $N=2$ theories.
The Higgs moduli correspond to the moduli of scalars
in the hypermultiplets
whereas the Coulomb branches correspond to the moduli of scalars
in the vector
multiplets.  Moreover there are no (F-type) mixtures
between hypermultiplets and vector multiplets and so the
two do not mix with each other.
Whereas the Higgs branches are easily computable using
classical Lagrangians of gauge theory, the same is not true
for the Coulomb branch which receives non-perturbative
point-like instanton corrections.  We are interested
in computing these corrections.  The simplification occurs
in type II theories on Calabi-Yau because the string coupling
constant is in a hypermultiplet
(see \ref\berk{N. Berkovits and W. Siegel, \nup B462 (1996) 213.}\
for a careful treatment).  Since the
 geometry
of the Coulomb branch is independent of hypermultiplet
vevs, we can take an arbitrary string coupling without
changing the answer \ref\stromi{A. Strominger, \nup 451 (1995) 96
}.
  This implies that if we compute
the tree level answer for Coulomb branch in string theory
it is the exact answer (this was in fact used in
\kacv \ref\fhsv{S. Ferrara, J. A. Harvey, A. Strominger and
C. Vafa, \plt 361 (1995) 59}).
We thus need to know the classical answer for the Coulomb
branch in type IIA string propagating
on the local geometry we have constructed.
However the classical answer on the type IIA side {\it does} receive
worldsheet instanton corrections.  In fact this
construction maps the contribution
of spacetime instantons of gauge theory to special
growth of the number of instantons of a two dimensional worldsheet
theory, as discussed in \kkv.  Mirror symmetry
comes to the rescue and results in summing up
the worldsheet instantons by giving a local mirror Calabi-Yau geometry
which gives an identical theory where we now consider type IIB strings
instead of type IIA.
In this case there are {\it no} worldsheet corrections and thus
the exact gauge theory answer can be read off from a classical
computation of a 2-dimensional theory.  This is thus our strategy:
Find the mirror of the geometry we have engineered
and then extract the exact quantum answers of the gauge system
by classical computations.
We thus see that mirror symmetry is a key fact allowing us
to extract exact answers.  We now turn to a review of certain
aspects of mirror symmetry.

\newsec{Intuitive Aspects of Mirror Symmetry and a Simple
Derivation of Exact Results}
In this section we will review some aspects of mirror
symmetry which provides an intuitive basis for the
results which will follow later in the paper.  Moreover in
this section we give a simple but less rigorous
derivation of some of our basic results
that will be rederived more rigorously using toric methods in later sections.
The cases we will derive the exact results for in this section include
the case of linear chain of $SU$ groups and $SU$ groups corresponding
to the affine $E_{6,7,8}$ quiver diagrams.

Consider a $d$-dimensional complex Calabi-Yau manifold $M$
and its mirror pair $W$.  This in particular means that for
$d$ odd type IIA(B) on $M$ is equivalent to type IIB(A) on $W$
where the role of K\"ahler deformations and complex deformations
get exchanged.  If $d$ is even, type IIA(B) on $M$ is equivalent
to type IIA(B) on $W$, again with the role of K\"ahler and complex
deformations exchanged.  Complex dimensions 1 and 2 are very special
because there are very few Calabi-Yau manifolds.
In dimension one there is only $T^2$ and in dimension 2
there is only $K3$ (apart from $T^4$ which has trivial holonomy).
This scarcity in low dimensions in particular leads to the fact
that $T^2$ and $K3$ are self-mirror.  The case of $T^2$ is very well
known and is a simple consequence of T-duality.  In the case
of $K3$ this is also a true but less trivial fact
\ref\mas{P. Aspinwall and D. R. Morrison,
{\it String theory on K3 surface},hep-th/9404151}.
Even though we will eventually be interested in the
case of complex dimension 3, aspects of $K3$ and its self-mirror
property play a crucial role in this section
and so we will now discuss it in a bit more detail.

As already discussed, we will only be interested in
a local model of $K3$ with singularities.  Let us recall
that the singularities
one encounters in
$K3$ are of ADE type.  Our local model will consist
of ALE space of ADE type and we are interested in constructing
the mirror.  Let us consider an $A_{n-1}$ ALE space.  This
can be described by a singular complex 2-manifold whose complex
structure is given by
\eqn\ale{xy+z^n=0,}
where $x,y$ and $z$ are complex numbers.  This space is singular
at the origin.  There are two ways this singularity can be remedied:
We can either deform the defining equation to make it less
singular or we can `blow up' the singularity.
The deformation which involves changing the complex structure
is given by
$$xy+\prod_{i=1}^n(z-a_i)=0,$$
with distinct $a_i$.  Up to a shift in $z$
there are $n-1$ physical parameters defining this deformation.
On the other hand we can resolve the
singularity by keeping the same defining
complex equation \ale\ but by `blowing up' the singularity, introducing
$n-1$ extra spheres which intersect one another
in the way dictated by the Dynkin diagram of $A_{n-1}$.  This
blow up is specified by $n-1$ complex parameters, corresponding
to the size and the $B$-field on each sphere.  In the blow up
one is varying the K\"ahler classes on the ALE space.
Mirror symmetry
in this case states that if we are interested in studying type
IIA(B) on this blow up space it is equivalent to studying
type IIA(B) on the complex deformed space exchanging
the $n-1$ complex parameters corresponding
to the complexified K\"ahler classes with $n-1$ complex
parameters describing deformation of the defining
equation.

In the applications we will consider it is also important to
consider the case where the local model for $K3$ involves an
elliptic fibration.  This is a well known subject mathematically
\ref\kod{K. Kodaira, Annals of Math., Vol. $\us{77}$, No. 3 (1963)}
\mir.  In particular the elliptic fibration over
the plane can develop ADE singularities as we change the
complex structure of the $K3$.  Again we can blow up the singularity
and we obtain new 2-cycles, a basis of which can
be taken to be ${\bf P}^1$'s which intersect according to the
Dynkin diagram of the ADE group, as was the case above.  The only new
ingredient in the elliptic case is that there is an extra special
2-cycle class, whose intersection with
the other cycles can be represented by an extra node
making the Dynkin diagram an {\it affine} Dynkin diagram.
If $s_i$ denote the Dynkin numbers associated with
each node of the Dynkin diagram, and if we denote
the $i$-th 2-cycle class by $C_i$ the 2-cycle class of the elliptic
fiber $E$ can be represented by
\eqn\aff{E=\sum s_i C_i}
Note that this is consistent with the fact that
$E\cdot E=0$.  The extra cycle
corresponding to the extra node on the affine Dynkin diagram
is of finite size even after all the other cycles have shrunk.
This follows from the fact that when all the other cycles
shrink the relation \aff\ implies that the size
of the extra 2-cycle corresponding to the affine
node is the same as the size of the elliptic fiber (recall
that the Dynkin number for the affine node is 1).

Mirror symmetry implies that the Kahler
deformation of the blow up is equivalent to complex deformation
of the mirror geometry.  The complex deformations of the mirror
has in turn another description which will prove useful
for us.  Consider type IIA string on a 2 dimensional
complex space with elliptic ADE singularity.  If we compactify further
on another $T^2$ we obtain an $N=4$ theory in $d=4$. From the viewpoint
of $N=1$ theory we have three adjoint chiral fields $X,Y,Z$ and a
superpotential
$$W={\rm Tr}[[X,Y],Z]$$
The Higgs branch of $N=4$ theory can be viewed as giving
vevs to the Cartan of $X$ or $Y$ or $Z$. In fact a $U(3)$
subgroup of the $SO(6)$ R-symmetry group rotates these fields
among each other.  We can identify $X$ with the blowing up
of the elliptic ADE singularity, $Y$ with the deformation of
the singularity and $Z$ as giving Wilson lines to the $ADE$
gauge group on the compactified $T^2$.  Given the R-symmetry
we deduce that three deformations are equivalent
and give rise to the same moduli space.
Thus we conclude, in particular, that
{\it the moduli space of blowups of elliptic ADE
singularities of complex surface is (mirror to) the moduli space of flat ADE
connections on a} $T^2$.  This result will be important later when
we discuss S-dualities that arise in field theories we study.
We will give an alternative derivation of this fact in section 8.

\subsec{Base Geometry vs. Fiber Geometry}
So far we have discussed mirror symmetry in complex dimension 2.
However for the purposes of the present paper
we are actually interested in the case of complex dimension 3.
The two are not unrelated, when we recall that we are interested
in fibering an A-type singularity over some collection of ${\bf P}^1$'s.
So roughly speaking all we have to do is to {\it also apply mirror
symmetry to the base as well}.  However, as we have discussed
before, the interesting class of configurations of the base also correspond
to when we have base ${\bf P}^1$'s which intersect according
to the ADE or affine ADE Dynkin diagrams. But we have already
discussed how these also arise in the complex 2-dimensional case.
 So {\it the mirror to both the base and the fiber
geometry will involve aspects of two dimensional mirror symmetry}
already discussed.  The only non-trivial data is how
a particular configuration of fiber ${\bf P}^1$'s
over the base ${\bf P}^1$'s is translated to mixing these two
mirror symmetry transformations.  We will now try to develop this
intuitively to arrive at a heuristic derivation
of some of the results which we will derive more rigorously later.

\subsec{An Intuitive  Derivation of Linear Chain
of $SU$ Groups}
Let us consider the case of engineering of a linear chain
of $SU$ gauge groups arranged along a line, with bi-fundamental
matter between the adjacent groups.  Let us suppose we have
$m$ gauge groups.  This means in particular that the
base geometry is $A_{m}$.  As discussed before
the mirror of this base geometry is given by
\eqn\lch{P_{m+1}(z)+uv=0}
where $P_{m+1}(z)$ denotes a polynomial
of degree $m+1$ in $z$ which is mirror to blowing
up the $A_m$ singularity.  The fiber geometry will be
a combination of $A_{n-1}$'s where $n$ varies over
each ${\bf P}^1$ in the base depending on the arrangement
of the $SU$ groups along the linear chain.  If we
denote the fiber variable $w$ (which together with $u,v,z$ and an equation
give a threefold), for each
of the $SU(n)$ factors in the fiber we expect to have
a polynomial $P_n(w)$ of degree $n$ in $w$.  This should
clearly be correlated with the coefficients in \lch ,
because the blowing up of the base geometry is mirrored
to complex deformation of the equation.  Suppose the
first group along the chain is $SU(n_1)$.  Then we should
see this group when we blow up the base only once, which is
mirror to
$$P_{m+1}(z)=z^{m+1}+a z^m,$$
where the $A_m$ singularity has been reduced to
$A_m\rightarrow A_{m-1}$.
At this point we should be able to see the fiber mirror because
we have blown up the base once and $SU(n_1)$ is the singularity
supported on the first blowup. This implies, applying
the mirror symmetry now to the fiber, that the coefficient
$a$ in the above equation should be a polynomial of degree $n_1$ in
$w$, i.e. we have for the defining equation of the threefold
$$z^{m+1}+P_{n_1}(w) z^m +uv=0.$$
Now we introduce the next blow up in the base which changes
the above equation to
$$z^{m+1}+P_{n_1}(w)z^m+b z^{m-1}+uv =0.$$
The last ${\bf P}^1$ that we have blown up
is reflected in the coefficient of the smallest power
of $z$ being non-zero.  Thus just as before, we now expect
$b$ to be a function $P_{n_2}(w)$ of degree $n_2$ in $w$.
Continuing this reasoning we will end up with the local
model for the threefold
$$\sum_{k=1}^{m+1}z^k P_{n_{m+1-k}}(w)+uv=0,$$
where we have put $P_{n_0}(w)\equiv 1$.
Note that the case of one gauge group (i.e. $m=1$) was
already considered in \klmvw\kkv\
which agrees with the above result.  Moreover it was noted
in \klmvw\ that one can use the T-duality which relates
$C^*$ fibrations in type IIB with NS 5-branes of type IIA
(and vice versa) \oov\ to show that this
is equivalent to considering an NS 5-brane of type IIA
whose worldvolume is $\Sigma \times R^4$ and where $\Sigma$ in this case
is a Riemann surface with equation
$$\Sigma : \qquad \sum_{k=1}^{m+1}z^k P_{n_{m+1-k}}(w)=0$$
carved out of the $(w,z)$ space.  Moreover it was
noted in \klmvw\ that the relevant metric on $\Sigma$
is provided by the SW  meromorphic 1-form on $\Sigma$.
This in particular shows that $\Sigma$ is non-compact.
Recently this result of \klmvw\ for one gauge group
 was rederived from the imbedding of
type IIA branes in M-theory in \witex\
and extended to the case of linear chain (with arbitrary
number of fundamentals)\foot{For an extension of these methods
to other gauge groups, see \witexf.}.  That result agrees with what
we have found above.  Later on in this paper we will generalize
our derivation to the linear chain of $A$-groups and in addition with
 arbitrary number of fundamentals
for each group and we fully recover the results of \witex\
from perturbative symmetries of strings.

\subsec{An Intuitive Derivation for affine E as the Base}
So far we only considered linear chains.  As discussed before
an interesting case involves the configuration of the $A$-groups
arranged along the affine ADE Dynkin diagrams, where the
rank of the $SU$ gauge group is proportional (with fixed proportionality)
to the Dynkin number of the corresponding node.  In this
case we get $N=2$ superconformal theories.  Here we show
how the general result for the Coulomb
branch of $A$-groups arranged according to the affine $E$ as the base
geometry can be obtained.

Just as in the linear chain considered
above we first need to know the mirror for the base geometry.
In particular we need a complex geometry whose deformation
is mirror to blowing up elliptic $E$-singularities.
To do this we recall that there are three special
constructions of $K3$ given by orbifolds which give
rise to elliptic $E_6$, $E_7$ and $E_8$ singularities:
$$K3={T^2\times T^2\over Z_3} \qquad E_6 \quad {\rm singularity},$$
$$K3={T^2\times T^2\over Z_4} \qquad E_7 \quad {\rm singularity},$$
$$K3={T^2\times T^2\over Z_6} \qquad E_8 \quad {\rm singularity},$$
where the $T^2$'s are special, in that in the first and third
case above they correspond to hexagonal lattice, and in the second
case they correspond to square lattice
(these singularities were studied in the context of F-theory compactifications
in
\ref\muk{K. Dasgupta and S. Mukhi, \plt 385 (1996) 125}
generalizing the work
\ref\sen{A. Sen, \nup 475 (1996) 562}).
We are interested in finding the mirror for these constructions.
Actually we will only be interested in the limit where one of the $T^2$'s
is replaced by ${\bf C}$, the complex plane, where an isolated
elliptic singularity appears.

 Let us first consider the $E_6$ case.  In this case the mirror is
given by the LG theory (modded out by an overall
$Z_3$) with the superpotential (see
\ref\va{C. Vafa, {\it Topological mirrors and quantum rings},
in {\it Essays on mirror manifolds}, edited by S.-T. Yau, International Press
1992.})
$$W=x^3+y^3+z^3+axyz+x'^3+y'^3+z'^3+a' x'y'z' +{\rm deformations}.$$
Here the unprimed variables denote the mirror to one torus and
the primed variables the mirror to the other.  Also
 the deformation monomials are of total degree 3 and mix up the unprimed
and primed variables.   There are three fixed points on each
torus which get identified with monomials $x,y,z$ and $x',y',z'$
and blowing up various fixed points, is mirror to choosing the combination of
corresponding monomials as deformation of the above potential.
Recalling that we are interested in the limit where the primed
torus becomes infinitely big (corresponding to sending $a'\rightarrow \infty$)
and concentrating on one fixed point on the ${\bf C}$ plane, corresponding
say to the variable $x'$, the deformation monomials will be of the form
$$x'^3,x'^2(x,y,z),x'(x^2,y^2,z^2).$$
Thus going to the patch where $x'=1$ and ignoring $y'$ and $z'$
which play no role in the above deformations we find that the
relevant deformation is given by the geometry
$$x^3+y^3+z^3+axyz+(b x^2+c y^2 +d z^2)+ (e x+f y +g z) +h =0.$$
Thus we have found the mirror
to blowing up elliptic $E_6$ singularity, where
the monomials $1,x,x^2$ correspond to the  blown
up ${\bf P}^1$'s of one fixed point on the $T^2$
and correspond to one edge of affine $E_6$ Dynkin diagram
starting from the trivalent vertex
of affine $E_6$, and similarly for $1,y,y^2$
and $1,z,z^2$.  Now we consider the fiber in addition to this
which will correlate with these monomials just as in the
case of the linear chain with the fiber geometry.
 Basically
we apply the idea of incorporating the fiber
in the linear chain case to each straight edge
of affine Dynkin diagram.
We find the local threefold
is now given by
$$\eqalign{0=\ &x^3+y^3+z^3+a xyz+P_{3k}(w)+\cr
&\sum_{i=1}^2 P^x_{i\cdot k}(w)x^{3-i}
+\sum_{i=1}^2 P^y_{i\cdot k}(w)y^{3-i}
+\sum_{i=1}^2 P^z_{i\cdot k}(w)z^{3-i}}$$
(it is also easy to write the geometry for the more
general case where the $\beta$-function is not zero, by choosing
polynomials of different degrees than those considered above,
just as in the linear chain case).
The generalization to the case where the base geometry
is affine $E_7$ or $E_8$ is straight-forward, where we start
with the elliptic curve $y^2+x^4+z^4$ for the $E_7$
case and $y^2+x^3+z^6$ for the $E_8$ case.   The
connection with blowing up of the $T^2\times C/Z_4$ and
$T^2\times C/Z_6$ are similar to the previous case.  In particular
for $E_7$ the monomials $x,z$ correspond to blowing up the
fixed point of order 4 whereas $y$ corresponds to blowing
up the fixed point of order 2.  In the case of $E_8$,
$z$ corresponds to the fixed point of order $6$, $x$
corresponds to the fixed point of order $3$ and $y$ corresponds
to the fixed point of order $2$. It is also easy to include
the fiber geometry as in the case of linear chain.   For the $E_7$ case with
vanishing $\beta$-function we have the mirror geometry
$$x^4+z^4+y^2+a xyz +\sum_{i=1}^3 P^x_{i\cdot k}(w) x^{4-i}+
\sum_{i=1}^3 P^y_{i\cdot k}(w) z^{4-i}
+ P^z_{2 k}(w) y +P_{4k}(w)=0$$
and for the $E_8$ case we obtain
$$x^3+y^2+z^6+a xyz +\sum_{i=1}^2 P^x_{2i\cdot k}(w) x^{3-i}+
 P^y_{3 k}(w) y+
\sum_{i=1}^5 P^z_{i\cdot k}(w) z^{6-i}+P_{6k}(w)=0.$$

Note that for these cases {\it we cannot reduce the data
solving the Coulomb branch of the} $N=2$ {\it system
to a Riemann surface}.  However we still have an equally
useful description of the Coulomb branch in terms of three
dimensional Calabi-Yau manifolds given above.
In particular one has a holomorphic 3-form whose
periods give the central terms in the $N=2$ SUSY algebra,
and one can read off the effective
coupling constants of the gauge theory from these periods, just
as would be the case for the $N=2$ solutions associated to
Riemann surfaces.

\newsec{Toric Geometry and Linear Sigma Models}

In the previous section we have seen how heuristic
applications of mirror symmetry goes a long way in giving
the type IIB geometry dual to a given type IIA geometry of
Calabi-Yau threefolds.  However for more general cases and
also to prove more rigorously the assertions of the previous
section, we need to recall in more detail
some of the machinery needed for this purpose
\ref\wili{E. Witten, \nup 403 (1993) 159}\ref\toli{P. Aspinwall, B. R. Greene
and D. R. Morrison,
\nup 416 (1994) 414}.
We first have to construct a 2 dimensional quantum field theory
which describes the propagation of type IIA strings in the
local model which we are interested in.  Next we have to use
this to construct the mirror geometry.  What we will do
now is to show how type IIA geometry can be summarized in terms
of {\it toric data} and how this can be used to construct the relevant
mirror.

In physical terms
the easiest way to construct the type IIA background
is in terms of linear sigma models \wili.
This involves considering
an $N=2$ gauge system with some matter which in the infrared
describes the conformal field theory corresponding
to the string propagation in a desired background.
For our purposes it suffices to consider the case with
gauge group $U(1)^r$, with $k$ matter fields $x_i$.  One
can also consider adding superpotential terms involving
the fields, and we shall need that for later applications.
There are also $r$ FI D-terms we can add to the theory,
one for each $U(1)$.
Let $q_i^a$ denote the $a$-th $U(1)$ charge of the $x_i$ field.
The condition that the theory has an extra R-symmetry (which
can thus flow to a non-trivial conformal theory) is that
\eqn\qsum{\sum_i q_i^a =0.}

To start with,
which is sufficient for some of the applications,
 we will consider a theory with no superpotential.
The vacuum configurations for this theory is described
by the gauge invariant fields.  It is well known
that this is the same as the manifold
\eqn\gvac{\IC^k/(\ICs)^r,}
where ${\bf C}^k$ corresponds to the complex values
for $x_i$ and where the $a$-the $\ICs$ action is given by
\eqn\cstar{x_i\rightarrow x_i \lambda^{q_i^a}.}
The manifold \gvac\ is the geometry
which the linear sigma model produces.   Note that
the complex dimension of this manifold is $d=k-r$.
This geometry is generically singular and putting the
FI D-terms into the Lagrangian has the effect
of resolving the singularity by blowing up the manifold.
Just as in the previous section, let us first concentrate
on the case of complex dimension 2.

\subsec{Type IIA on $A_{n-1}$ Background and Toric Geometry}
Let us consider our first concrete example.  We will
find the linear sigma model for $A_1$ singularity of $K3$.
For this purpose it is sufficient to consider the case of $U(1)$
gauge theory with three matter fields $x_i$ , whose charges are given by
\eqn\morio{
l^{(a)}\equiv(q_1^a,\dots,q_k^a)=(1,-2,1).}
The gauge invariant geometry (using \gvac) associated to this is
obtained by considering the generators
of gauge invariant (i.e. neutral) chiral fields
\eqn\invc{u=x_1^2x_2, \quad v=x_3^2x_2, \quad z=x_1x_2x_3.}
These are not independent, and there is one relation among them:
\eqn\reli{uv=z^2.}
So the geometry of the vacuum configuration \gvac\ in this
case is given by the $A_1$ ALE space.  Turning on the FI D-term
corresponds to blowing up the singularity.  To see this, note that
turning on the D-term corresponds to having the potential
\eqn\epot{V=(|x_1|^2+|x_3|^2-2 |x_2|^2-A)^2,}
where $A$ corresponds to the FI term. If we take $A>0$, $x_1,x_3$
cannot both be zero (in order to minimize $V$).  The coordinates
$(x_1,x_3)$ up to an overall rescaling (which gets identified
with the non-compact cotangent direction) can thus be viewed
as a ${\bf P}^1$ whose K\"ahler class is controlled by $A$.

The geometrical interpretation of the above field theory makes
it possible to use the powerful concept of toric geometry. This
is not really necessary for the simple example above, but will
be important for the more complicated cases, where the gauge
theory picture becomes quickly unmanageable, whereas the toric
methods proceed without trouble.

In toric geometry, the fields $x_i$ become homogeneous variables on
the quotient space $(\IC^k-U)/(\ICs)^{r}$, acted upon by
the $r$ $\ICs$ actions \cstar. $U$ is a subset of $\IC^k$,
defined by the $\ICs$ actions and a chosen ``triangulation''
and we will determine it in a moment. It generalizes the
point $x_i=0,\ \forall  i=1,\dots,n+1$, that is
removed in the case of  ordinary projective space $\IP^n$.

To find the gauge invariant fields we associate
to each field $x_i$
a vector $\nu_i=(\nu_{i,1},\dots,\nu_{i,d})$
in the standard lattice $\IZ^d$, such that the $\nu_i$
fulfill the following relations determined by the $\ICs$ actions
\eqn\vrel{
\sum_i \lm{a}{i}\nu_i = 0,\qquad \forall a=1,\dots,r \ .
}
Note that the dimension of the lattice is equal to the number of gauge
invariant generators, $d=k-r$. Furthermore, for any vector $k_j\in \IZ^d$,
$$
u(k)=\prod_i x_i^\mprod{\nu_i}{k},\qquad\quad
\mprod{\nu_i}{k}\equiv\sum_j\nu_{i,j}k_j,
$$
is a gauge invariant field
and we get therefore a  convenient representations of
the gauge invariant fields in terms of the integral
vectors $k_j$.

Since only positive powers of the fields $x_i$ should appear,
we make the further restriction, that if $N$ is the
lattice generated  by the vectors $\nu_i$ and if $M$
is its dual lattice, the allowed choice for $k$ lie in the
cone in $M$ defined by $\mprod{\nu_i}{k}\geq0$. Moreover,
to avoid
redundancy of the description, we restrict to a set of generators
$\ns\al\in M$ which generate all elements in this cone by positive
coefficients. The generators of invariant fields are then
$u_\al=\prod_i x_i^{\mprod{\nu_i}{\ns\al}}$.

Note that it follows from the anomaly freedom condition
\qsum\ that $\prod_i x_i$ is one
of the invariant monomials. This in turn implies\foot{Neglecting
subtleties related to torsion.} the existence
of a vector $h$ with
$\mprod{h}{\nu_i}=1\ \forall i$, that is the vertices $\nu_i$
lie in a hyperplane $H$ of $\IZ^d$. In geometrical terms
this corresponds to the condition, that the singularities
of the toric variety $V$ are sufficiently well behaved to
give rise to Calabi--Yau manifolds with first Chern class
$c_1=0$.
A convenient choice of coordinates is to take
$h=(1,0,0,\dots)$ and therefore $\nu_i=(1,*)$.

Applied to the above example defined by the charge vector \morio\
we get
$$
\nu_i=\pmatrix{1&-1\cr1&0\cr1&1},\quad
\ns\al=\pmatrix{1&-1\cr1&0\cr1&1},\quad
\mprod{\nu_i}{\ns\al}=\pmatrix{2&1&0\cr1&1&1\cr0&1&2}
$$
and thus $u_\al\equiv u(\ns\al)=(u,z,v)$ as in eq. \invc.

We still have to determine the disallowed set $U$.
For this we will need some more technical definitions;
however the final representation in terms of ``toric
diagrams'' will be very transparent and is all what
is needed to understand the following discussion.

The precise definition of the toric variety $V$
is in terms of a collection of cones $\sigma_\mu$,
bounded by rays
from the origin through the points in $\IZ^d$ defined
by the vertices $\nu_i$. This is shown in fig. 2.

\vskip 0.5cm
{\baselineskip=12pt \sl
\goodbreak\midinsert
\centerline{\epsfxsize 2.5truein\epsfbox{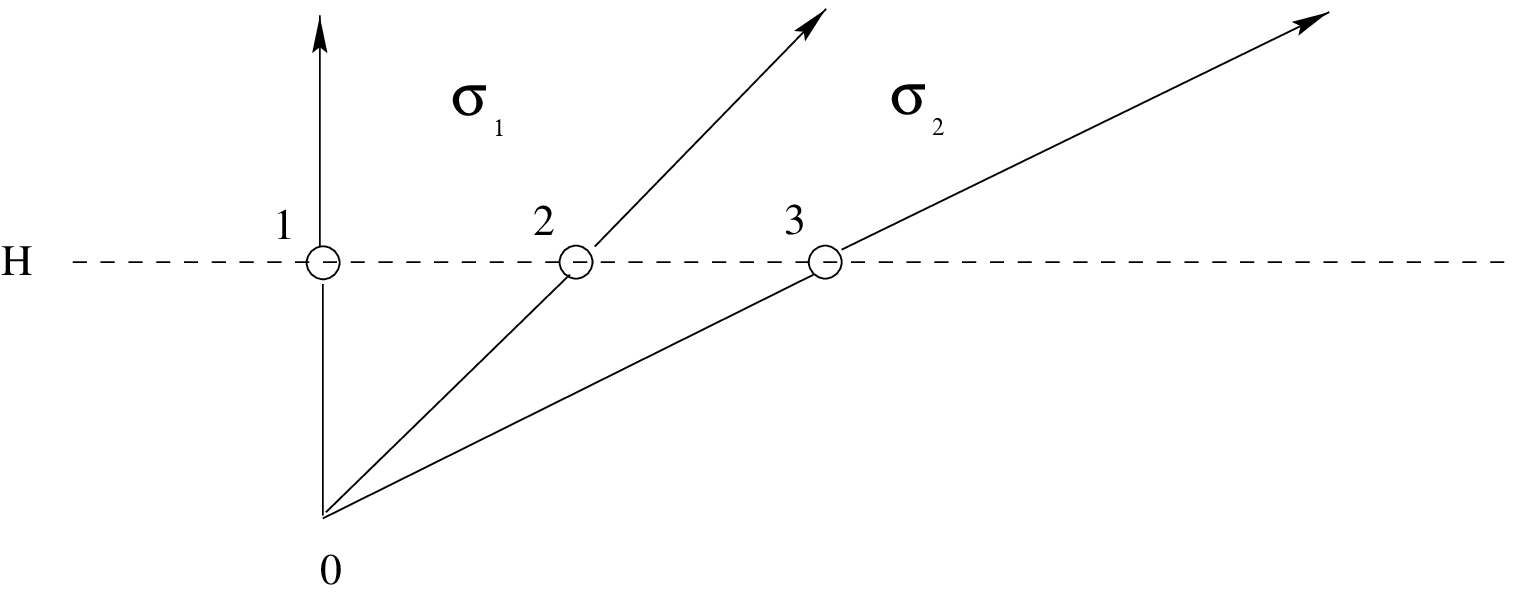}}
\leftskip 1pc\rightskip 1pc
\noindent{\ninepoint  \baselineskip=8pt {\bf Fig. 2}
}\endinsert}\ni
The disallowed set $U$  is now determined in the following way:
Elements in $U$ are defined by those subsets of vertices, which do not
lie together in a single cone $\sigma_\mu$:
$$
\{x_i=0,\ i\in\{i_\rho\}\} \in \ U\ \  {\rm if}
\ \ \{\nu_{i_\rho}\} \not\subset \sigma_\mu,\
\ \forall \sigma_\mu \ .
$$
In the example above, $U=\{x_1=x_3=0\}$, since the points $\nu_1,\nu_3$
do not lie in a cone (they are separated by the ray passing through
$\nu_2$).

{}From now on, we simplify the diagrams by suppressing the
direction normal to $H$. The toric diagram for the above configuration
looks than as in fig. 3.

\vskip 0.5cm
{\baselineskip=12pt \sl
\goodbreak\midinsert
\centerline{\epsfxsize 1.3truein\epsfbox{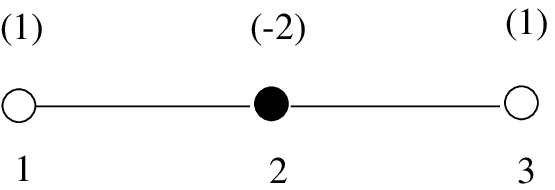}}
\leftskip 1pc\rightskip 1pc
\noindent{\ninepoint  \baselineskip=8pt {\bf Fig. 3}
}\endinsert}\ni
A nice property of the toric variety $V$ is that the
hyperplanes $D_i:\{x_i=0\}$, also called divisors,
generate the $d-1$ dimensional homology group and its dual,
the homology class of curves we are after.
In fact we are interested in the {\it
compact} part of the toric variety and the curve classes
$C_a$ contained in it.
The first rule to read the toric diagram
can be stated as: the divisor $x_i=0$ corresponding
to the node $\nu_i$ is compact if $\nu_i$ is
an interior node. In the example, the only interior node
is $\nu_2$, and $x_2=0$ can be readily seen to be
compact, since the potential
\epot\ implies $|x_1|^2+|x_3|^2=A$. Moreover the
$\IP^1$ with coordinates $x_1,x_3$ determined by this equation
is the only homology class of the compact divisor $D_2$.

Physically, the most important quantities of the type IIA geometry
are the intersections of the curve classes $C_a$ contained in $V$.
It is a standard calculation to determine these intersections
from the relations $\morio$ and $U$
(for a pedagogical review see \morI).
However in two
complex dimensions there is a nice short-cut, which, in the
presence of the fibration structure we use, will also be helpful for
the threefold case and provides a direct link between the
toric diagrams as in fig. 3 and Dynkin diagrams.
Similar observations about the appearance of Dynkin
diagrams in Calabi-Yau toric descriptions have been made in
\ref\canet{P. Candelas and A. Font,
{\it Duality between the webs of heterotic and type II vacua},
hep-th/9603170}
\ref\otp{P. Candelas, E. Perevalov and
G. Rajesh, {\it
Toric geometry and enhanced gauge symmetry of F theory /
                  heterotic vacua},
hep-th/9704097;\br
E. Perevalov and H. Skarke, {\it
Enhanced gauged symmetry in type II and F theory
                  compactifications: Dynkin diagrams from polyhedra},
hep-th/9704129}.
In fact there is the following simple way to read off the
curve classes and their intersections: {\it each curve class $C_a$
corresponds to a relation $\lm{a}{}$ between the vertices $\nu_i$.
Moreover the entries $\lm{a}{i}$ are the
intersections}\foot{More precisely, it can happen that
the intersections differ by a common normalization factor $N_a$.
We take care of these factors in the following.}\ {\it  $C_a\cdot D_i$}.
In particular, in two complex dimensions, the hypersurfaces $D_i$
are curves themselves and $C_a\cdot  D_i$ is the intersection
matrix for curves.

Note that the $k$ divisors $D_i$ are not independent but give rise
to $r$ different homology classes $K_a$, the Poincar\'e duals
of the curves $C_a$. If we choose a preferred basis
for the 2-cycles (thus fixing  certain linear combinations
of the $\lm{a}{}$), namely such that the volumes of the curve classes
$C_a$ generate the K\"ahler cone of $V$, the intersections
$C_a\cdot  D_i$ reproduce precisely the Cartan matrix of the
gauge system and part of the toric diagram agrees with the Dynkin diagram.
The charge vectors $\lm{a}{}$ in the basis dual to the K\"ahler
cone are called Mori vectors.

In the $SU(2)$ example above, the Dynkin diagram of $SU(2)$
is given by the middle node $\nu_2$ and we have also indicated
the intersections of the single curve class $C=D_2$ with
the non-compact divisors divisors $D_1,D_3$, (1), and
with itself, (-2).

Now we consider the generalization of this to $A_{n-1}$
ALE space.  We consider a $U(1)^{n-1}$ theory with $n+1$
fields, with charges given by
\eqn\mvsun{\eqalign{
\lm{1}{} &=(1,-2,1,0,0,0,\ldots ,0),\cr
\lm{2}{} &=(0,1,-2,1,0,0,\ldots ,0),\cr
\lm{3}{} &=(0,0,1,-2,1,0,\ldots ,0),\cr
&  \qquad \vdots \cr
\lm{n-1}{}&=(0,0,0,0, \ldots,  1,-2,1).}
}
It is not too difficult to read off the geometry associated
to this, just as in the $A_1$ case.  In particular the generators
of gauge invariant chiral fields are
$$\eqalign{u&=x_1^nx_2^{n-1}x_3^{n-2}\ldots x_{n+1}^0,\cr
v&=x_1^0 x_2^1x_3^2 \ldots x_{n+1}^n,\cr
z&=x_1x_2x_3 \ldots x_n,}$$
with the single relation
\eqn\relii{uv=z^n,}
which defines the background geometry corresponding to $A_{n-1}$
ALE space. Turning on the $n-1$ FI D-terms corresponds
to blowing up the ALE space and introduces $n-1$ K\"ahler classes.
Even though it is possible to study the linear sigma model
phases and see the geometry of the resolved space, this becomes
increasingly difficult.  In fact it is precisely to answer
such questions that toric geometry is useful.  So let us see how
this appears for the present example.

The toric data are now given by
\eqn\vertsun{
\nu_i=\pmatrix{1&0\cr1&1\cr . & .\cr . & .\cr . & .\cr1&n},\quad
\ns\al=\pmatrix{1&0\cr n&-1\cr0&1}, \quad
\mprod{\nu_i}{\ns\al}=\pmatrix{
1&n&0\cr
1&n-1&1\cr . & . & .\cr . & . & .\cr . & . & .\cr
1&0&n,}
}
summarized in the toric diagram fig. 4

\vskip 0.5cm
{\baselineskip=12pt \sl
\goodbreak\midinsert
\centerline{\epsfxsize 2.6truein\epsfbox{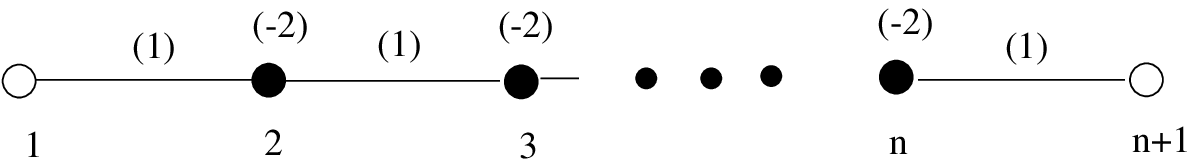}}
\leftskip 1pc\rightskip 1pc
\noindent{\ninepoint  \baselineskip=8pt {\bf Fig. 4}
}\endinsert}\ni
The Dynkin diagram of $SU(n)$ ,
associated to the curve classes $C_a$ in the
compact part of $V$, is plainly visible as the chain of black
dots. We have also indicated the intersections of the 2-spheres
contained in $D_i$, namely self-intersections $(-2)$ and intersections
$(1)$ with the next neighbor. Note that they are given precisely by
the entries  of the charge vectors \mvsun\ and moreover
agree with the entries of the Cartan matrix of $SU(n)$.

\subsec{Local Mirror Symmetry}
The data we used to describe the variety $V$, are the $k$
vertices $\nu_i$ spanning a polyhedron $\Delta$, the $r$ relations $\lm{a}{}$
fulfilled by them and the $k-r$ vectors $\ns\al$, that defined
the generators of gauge invariant fields $u_\al$. Note that the
vertices $\ns\al$ define similarly a polyhedron $\dd$
in the dual lattice. Batyrev's
construction of the mirror geometry
\ref\bat{V. Batyrev, J. Alg. Geom. $\us{3}$ (1994) 493}
proceeds by exchanging the roles of the polyhedra $\Delta$ and $\dd$
(for an attempt to prove the mirror symmetry in
terms of Batyrev's construction see
\ref\morpl{D. R. Morrison and M. R. Plesser,
{\it Towards mirror symmetry as duality for two-dimensional
abelian gauge theories}, hep-th/9508107}).
More precisely, we consider a Calabi--Yau manifold $X$.
In the simplest case, $X$ is defined as a hypersurface
in $V$, described by a homogeneous polynomial in the
gauge invariant monomials $u_\al$:
$$
0=p(X)=\sum_\al b_\al u_\al \ =\  \sum_\al
b_\al \prod_i x_i^\mprod{\nu_i}{\ns\al}
\ = \ b_0\; x_1x_2\dots x_k \ +\  \ldots \ ,
$$
where $b_\al$ are some complex numbers parameterizing the
complex structure of $X$. Similarly the mirror polynomial,
exchanging the roles of $\nu_i$ and $\ns\al$ is given by
\eqn\mirpolo{
0=p(\xs)=\sum_i a_i y_i \ =\  \sum_i a_i \prod_\al \xt_\al^{\ \
\mprod{\nst\al}{\nu_i}}
\ = \ a_0\; \xt_1\xt_2\dots \xt_k \ +\  \ldots \ ,
}
where $\xt_\al$ are the homogeneous coordinates of $V^\star$,
$\nst\al$ are vertices in the convex hull spanned by the
vertices $\ns\al$ and $a_i$ parameterize the complex structure
of $\xs$. Moreover $y_i$ are monomials in the variables $\xt_i$
invariant under the $\ICs$ actions $\xt_\al\to \xt_\al
\lambda^{\tilde{q}^a_\al}$,
which descend from relations $\sum \tilde{l}^{(a)}_\al\nst\al$
fulfilled by the dual vertices $\nst\al$.  The statement of mirror
symmetry is that the geometry of K\"ahler variations of the space
$X$ is captured by the complex deformations of the space defined by \mirpolo .
  Moreover there is a precise procedure to read off the instanton
corrections of the original space, in terms of ``variations of
hodge structure'' of the mirror geometry given by \mirpolo\
(in terms of specific period integrals).

The charge vectors $\lm{a}{}$ of $X$ imply the following
relations\foot{The relation \reli\ fulfilled
by the gauge invariant fields of the original manifold $X$
is a relation of this kind associated to the Mori generators of $\xs$,
$\tilde{l}^{(a)}_\al$.}
between the gauge invariant coordinates $y_i$ of
$\xs$:
\eqn\lmm{
\prod_i y_i^{q^a_i}=\prod_\al
\xt_\al^{\ {\mprod{\nst\al}{\sum_i \lm{a}{i}\nu_i}} }
=1 \qquad \forall a
}
Note that these equations can be studied for sets of relations
$\lm{a}{}$ independently of an embedding in a larger system. In particular,
as in \kkv,  consider the hyperplane given by
\eqn\mirg{p(\xs)=\sum a_i y_i=0,}
where $i$ runs only over the vertices $\nu_i$ describing the
local geometry of the gauge system. This equation is homogeneous,
and one can eliminate an overall scale from \mirg\
and so we end up with $k-r-2$ dimensional space as the mirror.
This can be two lower than the dimension expected in generic applications
in the compact cases as was in the $A_{n-1}$ case above.

This reduction of dimension (and the generalization to the complete
intersection case) can be understood in the following way.
Suppose we start with a (possibly non-compact) Calabi--Yau $X$ described
by $\hat{k}$ vertices
$\hat{\nu}_i$ and $\hat{r}$ relations $\hat{l}^{(a)}$. The dimension
of $X$ would be $\hat{d}=\hat{k}-\hat{r}$ without a superpotential
and $\hat{d}=\hat{k}-\hat{r}-2$ with a superpotential $p(X)$,
one less from the equation and one less because of \qsum.
Now divide the vertices $\hat{\nu}_i$ into two sets, the first,
$\Delta^0=\{\nu_i\}$, containing the $k$ vertices
describing the local geometry of the gauge system and the second one,
$\Delta^{0\ \prime}=\{\nu_i^\prime\}$ containing the rest. Similarly we divide
the relations $\hat{l}^{(a)}$ into two sets, according to whether or
not they involve elements of $\{\nu_i\}$. Let $r$ be the number
of relations $\lm{a}{}$ involving some of the $\{\nu_i\}$.

Two situations can arise: i) the $d$ dimensional local geometry
describing the gauge system is constrained, that is the
singularity exists only on the hypersurface $p(X)=0$.
In this case, $d=k-r-2$ and  the mirror geometry is of the
same dimension. This happens e.g. for a $D_n$ singularity
discussed in a later section.
ii) the $d$ dimensional local geometry
describing the gauge system is unconstrained. In this case
we have $d=k-r$, a case without a superpotential.
However the mirror geometry {\it is} constrained by $p(\xs)$,
giving a mirror of dimension $d=k-r-2=d-2$. In particular,
this happens for $A_n$ singularities where one obtains
Riemann surfaces as the mirror geometry of a threefold.
In such cases, one can relate the type IIB theory
to a $d$-fold geometry by noting that
adding two more variables to the equation
which appear quadratically, $p(\xs )=0\rightarrow
p(\xs)+uv=0$ does not affect the period integrals
and so we can view the mirror geometry as this local $d$-fold
(the trick of adding
quadratic variables to describe the geometry is
familiar from the study of LG models
\ref\gvw{B. Greene, C. Vafa and N. Warner, \nup 324 (1998) 371}
\ref\Mart{E. Martinec, {\it Criticality, Catastrophe
and Compactifications}, V.G. Knizhnik memorial volume, 1989.}).

Sometimes simplifications can occur in describing the mirror.
In particular if there are variables which appear only linearly,
they serve as `auxiliary' fields and can be eliminated
by setting to zero the variation of the polynomial with respect
to them, without affecting the period integrals of the mirror.
In particular if we have several variables $x_0^i$ which appear
linearly in the polynomial\foot{Note that the existence of the
hyperplane $H$ implies that there is always one variable which
appears only linearly, related to the hypersurface constraint
\mirpolo.},
$p(\xs)=\sum x_0^iG_i$, the mirror geometry can be viewed
as corresponding to the complete intersection
$G_i=0\ \ \forall i$.

Let us see how this mirror symmetry works in the case of $A_{n-1}$ which
we have constructed above.  In this case we have $y_1,...,y_{n+1}$
as the space of $y's$ subject to the relation \lmm
$$y_i y_{i+2}=y_{i+1}^2.$$
Choosing the homogeneous factor so that $y_1\rightarrow 1$ we can solve
the above relations\foot{Or simply using
$y_i=\prod \xt_\al^{\mprod{\nu_i}{\ns\al}}$
with $\mprod{\nu_i}{\ns\al}$ given in \vertsun\ and
$\xt_\al=(x_0,y,s\equiv1)$.} and obtain
$$(y_1,y_2,\ldots , y_{n+1})=(1,y,y^2,\ldots, y^n),$$
where we have defined $y=y_2$.  And thus the mirror geometry is
$$\sum_{i=1}^{n} a_i y^i= uv$$
where we have introduced the auxiliary variables $u,v$ to
make contact with 2-fold geometry.  The result is
 as expected, namely the K\"ahler deformations
of the $A_{n-1}$ singularity has been changed to deformation
of the same singularity (the self-mirror property of $A_{n-1}$
ALE space already discussed in section 3).

\subsec{The Threefold Case}
The type IIA compactification on the $A_{n-1}$
geometry described in the previous
section develops an enhanced $SU(n)$ gauge symmetry in six dimensions.
As discussed previously, to get an $N=2$ theory in four
dimensions we have to compactify
further on a one complex dimensional space which should have
no 1-cycles to avoid adjoint matter - that is a collection
of 2-spheres. Therefore we have to consider {\it base geometries that are
precisely of the same type as the fiber geometries}.

This makes the discussion of the threefold geometry simple.
As before let us start with the simplest case $A_1$, a single
2-sphere. However we have now two such $\IP^1$'s, one for the fiber and
one for the base. Moreover, instead of taking only the
naive product of the two $\IP^1$'s, we can take instead
non-trivial fibration of the first $\IP^1$ over the second
one (see \kkv\ for a detailed treatment of the mirror
of such cases that we review below).

These $\IP^1$ bundles over $\IP^1$ are classified by a single
integer $n$ and are called Hirzebruch surfaces, denoted $\IF_n$.
Let us describe them in the notation we introduced in the previous
section. The geometry of the $A_1$ singularity with a single blow up
sphere we considered in detail, was conveniently summarized in
the toric diagram fig. 3. For two $\IP^1$'s, one for the
fiber and one for the base, we will have to combine
two of these geometries. The only non-trivial question is
how they are connected, in other words, to specify the
fibration. Let us continue to reserve
the horizontal direction in the diagrams to denote the
fiber geometry whereas we use now the vertical direction
to draw the geometry for the base. The result is shown
in  fig 5.

\vskip 0.5cm
{\baselineskip=12pt \sl
\goodbreak\midinsert
\centerline{\epsfxsize 3.2truein\epsfbox{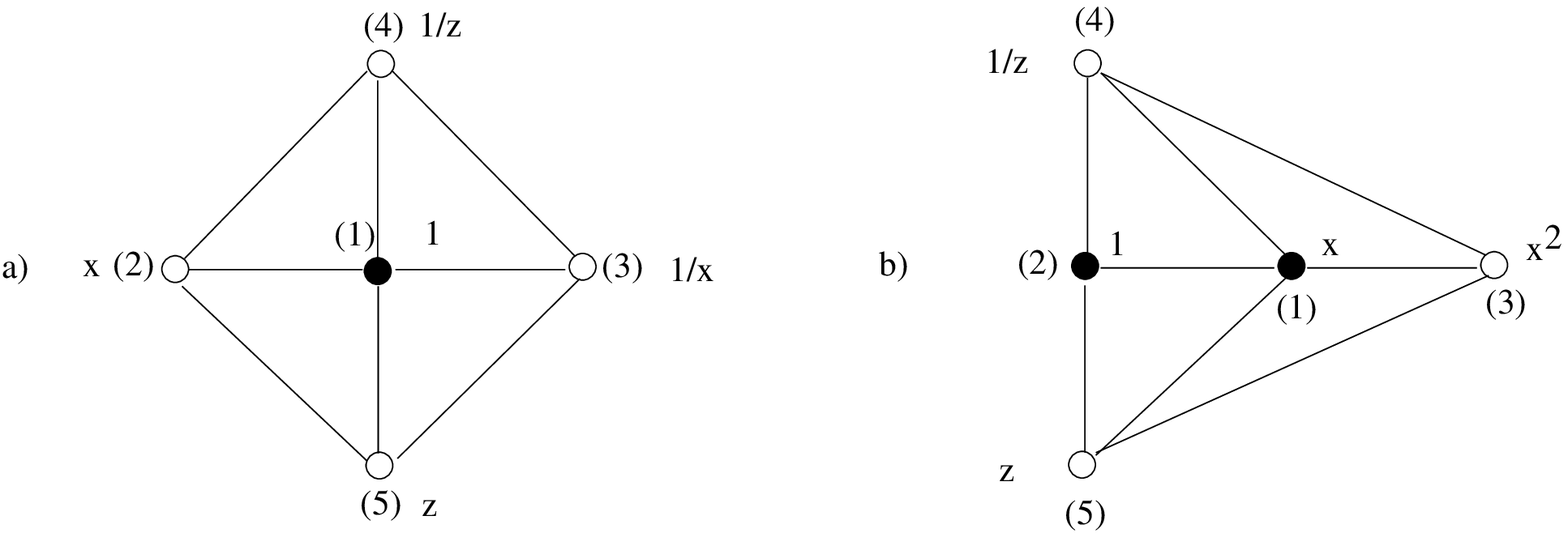}}
\leftskip 1pc\rightskip 1pc
\noindent{\ninepoint  \baselineskip=8pt {\bf Fig. 5:}
Local threefold geometries: a) $\IF_0: \IP^1\times \IP^1$\ b) $\IF_2$.
}\endinsert}\ni
Note that the fiber (base) $\IP^1$ corresponds to simply omitting
the points $\nu_4,\nu_5$ ($\nu_2,\nu_3$).
Not surprisingly, the two geometries in fig. 5 differ
in the fibration, the first one corresponding to the trivial product
$\IF_0:\; \IP^1\times\IP^1$ whereas the second geometry  has a non-trivial
fibration and is an $\IF_2$ surface.
To see this, let us write down the vertices and charge vectors:
\eqn\exfn{{\ninepoint{\eqalign{
\IF_0&:\nu_i=\pmatrix{
0&0\cr-1&0\cr1&0\cr0&-1\cr0&1\cr},\qquad
\lm{a}{}=\pmatrix{-2&1&1&0&0\cr-2&0&0&1&1},\cr
\IF_2&:\nu_i=\pmatrix{
0&0\cr1&0\cr-1&0\cr-1&1\cr-1&-1\cr},\qquad
\lm{a}{}=\pmatrix{-2&1&1&0&0\cr0&0&-2&1&1}.}}}
}
Recalling that the charge vectors $\lm{a}{i}=q^a_i$ define
the $\ICs$ actions $x_i\to\lambda^{q^a_i} x_i$, we see that in
the first case the projective actions of the $\IP^1$ factors
are independent, $(x_1,x_2,x_3,x_4,x_5)\to
(\la^{-2}\mu^{-2} x_1,\la x_2,\la x_3,\mu x_4,\mu x_5)$,
whereas in the second case the coordinates of the fiber
$\IP^1$, namely $(x_2,x_3)$, transform non-trivially under
rescalings of the coordinates $(x_4,x_5)$ of the base $\IP^1$,
$(x_1,x_2,x_3,x_4,x_5)\to
(\la^{-2} x_1,\la x_2,\la\mu^{-2} x_3,\mu x_4,\mu x_5)$.
We have also indicated in
fig. 5 the variables that solve \lmm\ and appear in
the hypersurface constraint $p(\xs)$.

Note that the first coordinate, $x_1$, is necessary to satisfy the
anomaly cancellation \qsum. It corresponds
to the non-trivial canonical bundle of the Hirzebruch
surfaces $\IF_n$. To ensure the Calabi--Yau condition $c_1=0$ we
have to consider the 3 complex dimensional total space.

In the large base limit, which is the relevant one for the
weakly coupled field theory limit (see \cov), the difference
between the two fibrations $\IF_0$ and $\IF_2$
is actually irrelevant. It is only the
stringy strong coupling behavior, corresponding
to small base, in which they differ. We will
discuss these points as we go along with the solution of
the more general theories.

Let us finally sketch the appearance of matter. As explained
already, matter arises from extra singularities, localized
above special points on the base geometry. Geometrically this
corresponds to introducing extra $\IP^1$'s, blowing up points
on the base. This is shown for the $SU(3)$ gauge theory
with $N_f=1$ matter in fig. 6 a).
\vskip 0.5cm
{\baselineskip=12pt \sl
\goodbreak\midinsert
\centerline{\epsfxsize 4.2truein\epsfbox{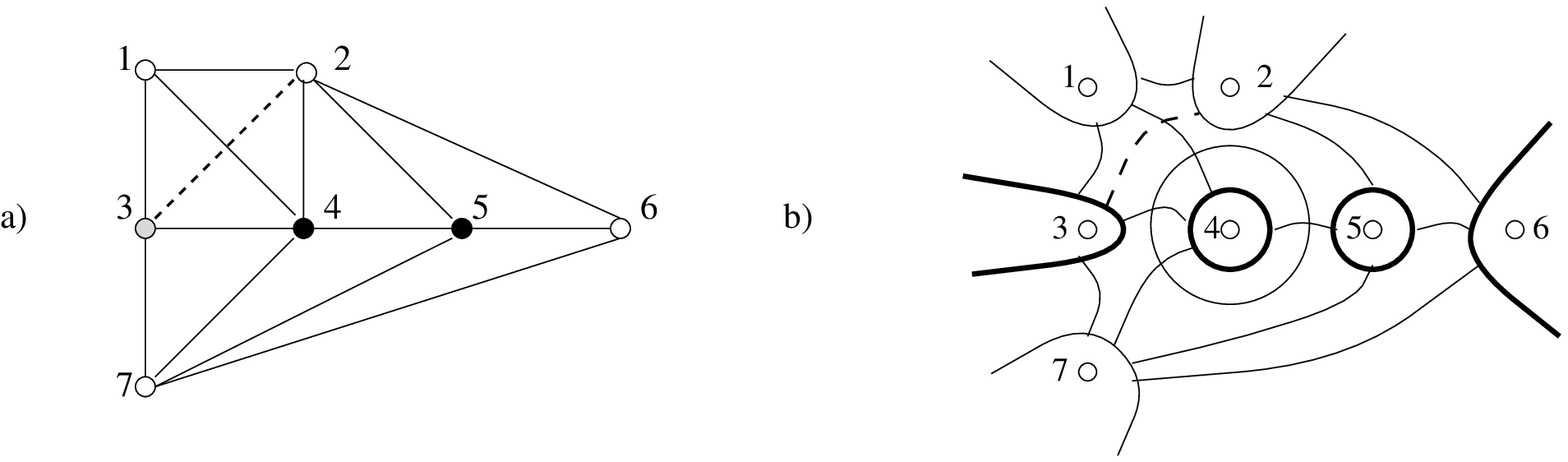}}
\leftskip 1pc\rightskip 1pc
\noindent{\ninepoint  \baselineskip=8pt {\bf Fig. 6:}
Geometry for the $SU(2)$ $N_f=1$ theory: a) toric polyhedron $\Delta$
of the type IIA geometry b) Riemann surface of the type IIB geometry.
}\endinsert}\ni
As in the previous example, the base is a simple $\IP^1$
factor represented by the three points $\nu_1,\nu_3,\nu_7$ on the
vertical line. The horizontal line with points
$\nu_3,\nu_4,\nu_5,\nu_6$ describes the case $n=3$
of eq. \vertsun, an $A_2$ singularity. Inspection of the relations
$\lm{a}{}$ as in \exfn\ identifies the compact divisors $\nu_4$
and $\nu_5$ as a blowup of $\IF_1$ and the
$\IP^1$ bundle $\IF_3$, respectively.
The blow up $\IP^1$ corresponding to the matter corresponds to the
extra point $\nu_2$.

The above geometry is also a simple example where it is possible
to choose two different partitions into cones, denoted by the
dashed line in the figure. In particular this means that the specification
of the vertices alone does not determine the geometry completely.
Drawing rays through the points $\nu_i$, as in  fig. 2,
is not sufficient to generate a valid collection of cones,
which is characterized by the property that the projection
of the faces to the hyperplane $H$ we draw, should yield a triangulation
of the polyhedron $\Delta$ defined by the points $\nu_i$.

Physically, a choice of triangulation corresponds to the fact
that the spectrum of light relevant BPS states can depend on
the region in moduli space one considers. A simple representation
of this fact is shown in fig. 6b) which displays the Riemann
surface $E$ representing
the mirror geometry of the toric
geometry in fig. 6a). Each interior point of the polyhedron $\Delta$
corresponds to a non-trivial homology class of $E$ and moreover each link
of $\Delta$ to a non-trivial 1-cycle on $E$. Moduli
of the gauge theory are associated with periods along 1-cycles in the
compact part of $E$ whereas bare parameters as the gauge couplings
mass parameters arise from 1-cycles that wrap the non-compact legs
of $E$, related to the behavior ``at infinity''. Depending on the
region in moduli space, the period over the 1-cycles between the
points $\nu_1$ and $\nu_4$ may be smaller or bigger than the
one over the dashed cycle between the points $\nu_2$ and $\nu_3$,
Choosing the one that leads to the smaller value corresponds
to choosing the triangulation of $\Delta$.

\subsec{Mirror Map on Moduli Space and the Exact Solution}
Let us finally collect the necessary ingredients to determine
the exact solutions for the moduli dependent gauge couplings
from the geometry.

In the type IIA compactification, the volumes $V_a$ of the 2-spheres
combine together with the anti-symmetric tensor fields $B_a$
to form complex fields $t_a$ parameterizing
the K\"ahler moduli space. In the Mori basis we introduced previously,
the K\"ahler moduli $t_a$ correspond 1-1 to the charge vectors
$\lm{a}{}$, as is clear from their interpretation as FI parameters
in the linear sigma model. Mirror symmetry, now on the moduli space,
relates the K\"ahler moduli $t_a$ of the type IIA geometry $X$
to complex structure moduli $z_a$ of the mirror geometry $\xs$
$$
t_a=B_a+iV_a=\ov{1}{2\pi i} \ln z_a + \cx O (z_a) ,
$$
where the so-called algebraic coordinates $z_a$ are given by
\eqn\algc{
z_a=\pm \prod_{i=1}^{r}a_i^{\lm{a}{i}}
}
with the $a_i$ defined as in \mirpolo. The exact worldsheet instanton
corrected prepotential is then obtained from the period integrals
of the unique holomorphic $(d,0)$ form $\Omega$ on $\xs$,
in the general form given in \bat. Another expression which 
turns out to be useful in certain cases, is the logarithmic form:
\eqn\form{
\Omega = \ln( p(\xs) )\prod_{i=1}^{m} \ov{dy_i}{y_i} \ .
}
It is straightforward to check, that the period integrals
$\Pi_i=\int_{\gamma_i} \Omega$ of $\Omega$, where $\gamma_i$ is a basis of
non-trivial
homology $d$-cycles, fulfill the GKZ system of differential equations
\eqn\gkz{
\prod_{{\lm{i}{k}}>0} \big(\ov{\partial}{\partial a_i}\big)^{\lm{i}{k}} =
\prod_{{\lm{i}{k}}<0} \big(\ov{\partial}{\partial a_i}\big)^{-\lm{i}{k}} \ .
}
Moreover, as a consequence of the $\ICs$ actions,
the period integrals depend only on the combinations $z_i$ of the $a_{i}$
defined in \algc. This invariance can be similarly expressed in terms
of differential equations. They depend in addition on a choice of 
gauge for $\Omega$ which fixes the proportionality
factor that depends only on the complex parameters $a_i$.
The Calabi--Yau periods are then obtained by choosing 
linear combination of the solutions
to \gkz\ with leading behavior determined by the intersections. Specifically,
the perturbative one-loop prepotential of the four-dimensional $N=2$ gauge
theory,
defined by a choice of a root system $R_G$ corresponding to a gauge group $G$
and matter representations corresponding to weight vectors
$w \in W_G$, is given by
\eqn\ppI{
\cx F = \ov{i}{4\pi}\sum_{\al \in R_G} (a \cdot \al)^2 \ln \ov{(a \cdot \al)^2}
{\Lambda^2}
-\ov{i}{4\pi}\sum_{w \in W_G} (a \cdot w)^2 \ln \ov{(a \cdot w)^2}{\Lambda^2}
,}
where $a_i$ are the Coulomb moduli parameterizing the moduli
space of the $N=2$ theory. From the string point of view, the
perturbative couplings \ppI\ describe the intersections of the
type IIA geometry $X$. Once we have completed the geometrical engineering
of the appropriate local geometry of 2-spheres, such that the intersections
reproduce the perturbative piece \ppI, the exact solution is immediately
determined in terms of the Picard-Fuchs system \gkz\foot{For
details on Calabi--Yau techniques to determine the periods
from the GKZ system, see
\ref\hkty{
S. Hosono, A. Klemm, S. Theisen and S.T. Yau, \cmp 167 (1995) 301}
\ref\cyrefs{P. Aspinwall, B. R. Greene and  D. R. Morrison,
\nup 420 (1994) 184}.}.
Note the amazingly direct relation between the group theoretical data
$(\al\in R_G, w \in W_G)$ defining the perturbative field theory,
and its exact geometrical solution in terms of the differential
equations \gkz. {\it Essentially all we need is the direct correspondence
between the charge vectors $\lm{a}{}$ and $(\al,w)$ defining the
2d and 4d field theories, respectively.}  This is similar
to statements about the relation of Picard-Fuchs equations with the
matter representations made in the context of $N=2$ field theories
 in \ref\schn{J.M. Isidro, A. Mukherjee, J.P. Nunes and H.J. Schnitzer, {\it A
new derivation of the Picard-Fuchs equations for N=2 Seiberg-Witten theories},
hep-th/9609116;
{\it A note on the Picard-Fuchs equations for N=2 Seiberg Witten
Theories}, hep-th/9703176;
{\it On the Picard-Fuchs equations for massive N=2 Seiberg-
                  Witten theories}, hep-th/9704174}\ref\alish{M. Alishahiha,
{\it On Picard-Fuchs
equations of the SW models},
hep-th/9609157;
{\it Simple derivation of the Picard-Fuchs equations
for the Seiberg-Witten models}, hep-th/9703186}.

The periods $\Pi_i$ describe the exact special geometry of the Calabi--Yau
moduli
space. To get the rigid special geometry of the field theory moduli space
we have still to decouple gravity effects, as in \kklmv.
We will describe the
appropriate limit when we treat the general case of a product gauge group in
section
sect 5.2.

\newsec{Linear Chain of $\prod_i SU(k_i)$ with bi-Fundamental
Matter}

We will now describe the geometrical construction of the exact solution of
the linear chain of $SU$ groups with bi-fundamental
matter between adjacent groups. We begin with a simple case of
$SU(N+1)$, $N_f=1$ and gradually add the different building blocks needed to
describe the most general case.

\subsec{$SU(N+1)$ with $N_f=1$}
We start with the construction and solution of the model shown in fig. 6,
with the generalization that we consider general $SU(N+1)$ instead of
$SU(3)$. The type IIA geometry $X$ is therefore defined by the $n_\nu=N+5$
vertices $\nu_i$ in $\IR^2$:
\eqn\verti{\ninepoint
\nu_{1,0}=(1,0),\ \nu_{2,i}=(2,i),\ \nu_{3,k}=(3,k) \ ,}
with $i\in \{0,\dots,N+1\}$ and $k \in \{0,1\}$. To make contact with the
gauge theory we identify the classes $C_i$ of the 2-cycles and their
intersections. As explained previously, the classes $C_i$ are
in 1-1 correspondence with linear relations of the vertices
$\nu_i$, described by the charge vectors
\eqn\movi{
\eqalign{\ninepoint
\lm g{}   &= (1,-2,0^{N+1},1,0),\cr
\lm{c}{i} &= (0^{i},1,-2,1,0^{N-i},0^2),\ i=1,\dots,N\ ,\cr
\lm m{}  &= (0,-1,1,0^N,1,-1) \ ,}
}
where we introduce the following notation: parameters related to bare gauge
couplings will be denoted by a superscript $(g)$, and similarly bare masses
by $(m)$ and Coulomb parameters by $(c)$.

\subsubsec{Local Type IIA Geometry}
Let us describe the local geometry of the above toric variety  $X$ in some more
detail;
in particular we want to show that it contains the homology of 2-cycles with
the
appropriate intersections. The derivation
in the following paragraph
requires some more knowledge of
toric geometry which is not needed otherwise to follow the remaining
discussion.

The polyhedron $\Delta$ defined by $\verti$ describes a toric variety, whose
compact part is composed of a chain of $N$ rational ruled surfaces $E_i$,
$ i=1,\dots,N$. In the above model, the $E_i$ for $i>1$  are Hirzebruch
surfaces $\IF_{n_i}$
with $n_i=2i-1$, and the first one, $E_1$, is a blow up of $\IF_2$
along the intersection
of the section and the fiber. The blowup corresponds to the
additional vertex $\nu_{3,1}$.
The curve classes can be described as follows: each
$\IF_{n_i}$ has
two sections $s_i,\ t_i$ with intersections $s_i^2=-n_i,\ t_i^2=n_i,\ s_i \cdot
t_i = 0$
and a fiber $f_i$ class with $f_i\cdot s_i=1$. Since the compact divisors $E_i$
meet along
sections, there is only one independent class from the sections, which we can
choose to be the section of the first factor, $E_1$. In addition we have
$N$ fibers $f_i$ and the exceptional curve of the blow up, $u$, which supports
the
matter hypermultiplet. In summary we have $N+2$ curve classes $s_1,f_i,u$ with
the following intersections with the divisors $E_i$:
\eqn\isi{{\ninepoint\eqalign{
s_i\cdot E_i&=\cases{                        -2 & if $i=0$\cr
                         0 & otherwise.\cr},\
f_i\cdot E_j=\cases{    -2 & if $i=j$\cr
                         1 & if $|i-j|=1$\cr
                         0 & otherwise.\cr},\cr
u \cdot E_i&=\cases{     -1 & if $i=0$ \cr
                         1 & if $i=1$ \cr
                         0 & otherwise.\cr}}
\ \ }}
where we denoted by $E_0$ the non-compact divisor at one end of the
$A_{N}$ chain.
Note that these intersections are precisely the entries of the Mori vectors
\movi. Moreover, $\lm g{}$ and $\lm m{}$
describe the charges of the $SU(N+1)$ vector bosons and the highest weight
of the fundamental matter in a Cartan basis, respectively.

\subsubsec{Type IIB Mirror Geometry:}
The intersections \isi\ guarantee the correct leading behavior described
by the perturbative gauge couplings \ppI. To solve for the exact dependence
including instanton corrections, we apply now the
local mirror transformation \lmm\ to
the above type IIA geometry parametrized by the K\"ahler moduli
to obtain the type IIB geometry parametrized by complex deformations.
A solution is given by the monomials
$$
\ov{1}{z},s^{N+1},s^{N}w,\dots,w^{N+1},zs^{2N+2},zws^{2N+1} \ .
$$
and after setting $s=1$, using the $\ICs$ action
$(z,w,s) \to (\lambda^{-N-1}z,\lambda w,\lambda s)$, we obtain
the defining equation for the mirror manifold $\xs$,
a Riemann surface:
\eqn\pola{
p(\xs) = \ov{a_{1,0}}{z}+P_{N+1}(w)+z(a_{3,0}+a_{3,1}w)\ ,
\ \ P_{N+1}(w)=\sum_{n=0}^{N+1} a_{2,n}w^n \ .
}
The moduli $a_{i,j}$ are related to the vertex $\nu_{i,j}$ carrying
the same index. Note that although the curve \pola\ is appropriate
to describe the field theory, the  expression \form\ for the holomorphic
1-form, $\Omega=\ln(w)dz/z$ is not (yet). Only after taking the large base
space limit, which requires a shift of the $w$ variable to stay near
the singularity, the answer will agree with the field theory answer.
We discuss the precise limit together with the more general case in the next
section.

\subsec{The Linear Chain with bi-Fundamentals}

Let us consider now the general case of products of $SU(N)$ factors with matter
in
the bi-fundamental representations. In the case of a single $SU(N)$
factor, we had a single $\IP^1$ as the base geometry. For the generalization
to a product of $SU(N)$ factors we have simply to replace the single $\IP^1$
by an $A_M$ chain of $\IP^1$'s with nearest neighbors intersecting.
The toric diagram now looks like shown in fig. 7.
\vskip 0.5cm
{\baselineskip 12pt\sl
\goodbreak\midinsert
\centerline{\epsfxsize 1.65truein\epsfbox{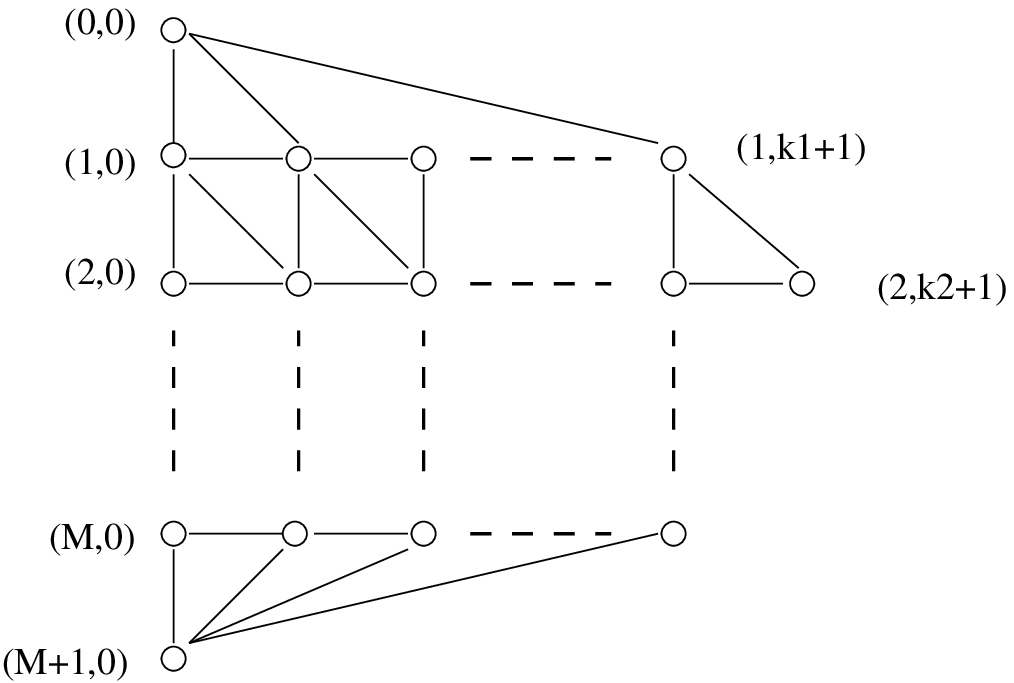}}\vskip0.5cm
\leftskip 1pc\rightskip 1pc\noindent{\ninepoint  \baselineskip=8pt {\bf Fig.
7:}
Polyhedron $\Delta$ for the product gauge group $\prod_{r=1}^M \; SU(k_i+1)$.
}
\endinsert}\ni
The vertical line at the left end describes an $A_M$ Dynkin diagram
associated with the base geometry, whereas the horizontal lines
describe the $SU$ factors from the fibers.

In fig. 7 we have implicitly assumed the convexity of
the polyhedron $\Delta$. The convexity assumption is needed for the validity
of the mirror description in terms of polyhedra.  This can
be easily seen to be equivalent to the appropriate high-energy behavior
of the product gauge theory: The matter content of a given $SU(k_r+1)$ factor
is $(k_{r-1}+1)+(k_{r+1}+1)$ fundamental representations. Asymptotic freedom
requires $2k_r \geq k_{r-1}+k_{r+1}$, in agreement with the convexity of
the polyhedron $\Delta$.

However also the non-convex toric diagrams have a valid physical
interpretation.
Assume we start from a convex toric diagram as in fig. 7,
but tune the moduli
of some of the vertices on the right side towards zero. This is an allowed
moduli and leads to some ${\bf P}^1$'s
becoming large.  Therefore the corresponding
gauge bosons and matter fields associated with D2
branes wrappings get very heavy. At low energies, the theory is described
effectively by states associated with the small 2-cycles, which might well be
described by the homology classes of a non-convex toric diagram. However
we should think of such a diagram as being part of a more complicated
theory, including the vertices required by convexity.

There are two types of
situations which may arise; the toric diagrams are depicted in fig. 8.
In the first case, completing the diagram effectively enlarges the rank
of the previously infrared free $SU$ factor. In this case the consistent
high energy behavior is restored by the presence of additional charged
vector bosons and their negative contribution to the beta function.
In the second, more interesting case, the rank of the gauge group stays the
same. The consistent high energy behavior arises from a non-trivial coupling
of the infrared free $SU$ factor to the other group factors.
\vskip 0.5cm
{\baselineskip 12pt\sl
\goodbreak\midinsert
\centerline{\epsfxsize 3.0truein\epsfbox{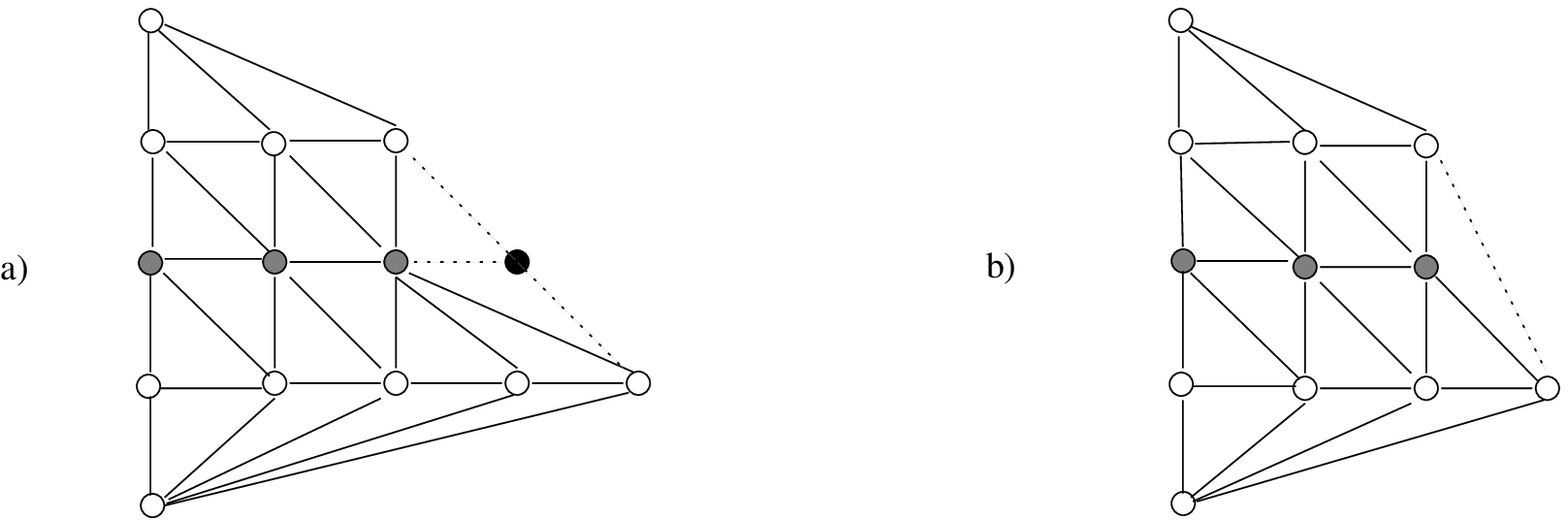}}\vskip0.5cm\leftskip 1pc
\rightskip 1pc\noindent{\ninepoint  \baselineskip=8pt {\bf Fig. 8:}
Toric completion of an infrared free gauge group factor (grey vertices)
a) convexity requires an additional {\it vertex} $\nu^\prime$ (black circle), \
b) the  completion of the convex
polyhedron requires only a new {\it link} between
the neighbored $SU$ factors (dotted line).  }
\endinsert}\ni
The vertices $\nu_i \in \IR^2$ of the polyhedron $\Delta$ can be read off
from fig. 7. For a product gauge group $\prod_{r=1}^M \; SU(k_r+1)$,
we have $n_\nu=2+2M+k$ vertices, where $k=\sum_rk_r$. There are
$n_{C_i}=2M-1+k$
independent classes of 2-spheres $C_i$, whose volumes describe the $k$ Coulomb
parameters $\zc_{r,i}$, $M-1$ bare masses $\zm_s$ and $M$ coupling
constants
$\zc_i$:
\eqn\aco{
\vbox{\offinterlineskip\tabskip=0pt\halign{\strut
$#$~\hfil &$#$~\hfil &$#$~\hfil &$#$~\hfil \cr
\zc_{r,i}&={a_{r,i-1}a_{r,i+1} \over a_{r,i}^2}
&,\ r=1,\dots M&,\ i=1,\dots,k_r\ , \cr
\zm_{r}&={a_{r,0}a_{r+1,1} \over a_{r+1,0}a_{r,1}}&,\ r=1,\dots,M-1\ , &\cr
\zg_{r}&={a_{r-1,0}a_{r+1,0} \over a_{r,0}^2}&,\ r=1\dots M \ .&\cr
}}}
The naive dimension of the mirror manifold is $n_\nu-n_{C_i}=3$.
However this is a case where we had no superpotential on the
type IIA side and we expect a reduction of dimension by two as explained
previously.
The Mori vectors $\lm{i}{}$ can be easily read off from  equations \aco,\
\algc.
The local mirror geometry, given as the solution of \lmm\ with the Mori vectors
defined as above, describes a Riemann surface $\xs$ given as a hypersurface:
$$
p(\xs)=a_{0,0}+\sum_{r=1}^M\; z^r P_r(w)+a_{M+1,0}z^{M+1},\qquad
P_r(w) = \sum_{l=0}^{k_r+1}a_{r,l}w^l.
$$
To reduce to the field theory limit, we shift $w$ by
a constant $\sim \ep^{-1}$ and send $\ep\to 0$. This
shift has to be accompanied by a corresponding limit in moduli
space to stay near the singular point of the $A_{k_r}$ singularity
described by $P_r(w)$. This identifies
\eqn\ftl{
a_{r,l}\sim \ep^{l-k_r-1}
}
as the correct limit. Furthermore $a_0, a_{M+1} \sim \ep^0$.
The algebraic coordinates $\aco$ scale in this limit as
as
$$
\zc_r \sim \zm_r \sim \ep^0,\qquad
\zg_r \sim \ep^{-b_r}\ ,$$
where $b_r=k_{r-1}+k_{r+1}-2k_{r}$ is the beta-function
coefficient of the $r$-th group factor.
For asymptotic free theories $b_r < 0 $ and the
limit $\ep \to 0 $ corresponds to taking a very large base, as
expected: From \algc\ the volume of the $r$-th 2-sphere
in the base scales as $V_r \sim b_r \ln \ep$.
Moreover, for
each configuration saturating the limit, $2k_{r+1} = k_r+k_{r+2}$,
we gain a new finite parameter $\zg_{r+1}$ in the field theory limit,
corresponding to an undetermined bare coupling $\tau_r$. Note that
this  parameter appears naturally as the Coulomb modulus of a
$SU$ factor from the base. This will be important when we
identify the S-duality group and its physical origin.

We complete the discussion of the chain model with an
explanation of the local type IIA geometry. Similarly as in the
case of a single
$SU(N)$ factor, the interior points of the polyhedron describe
the compact divisors $E^i_r$ of $X$. We have $i=1,\dots,k_r$ of such
exceptional divisors, ruled over the
$r$-th base $\IP_r^1$, which itself intersects transversally the
adjacent spheres $\IP^1_{r-1}$ and $\IP^1_{r+1}$ in the base.
The divisors $E^i_r$ are
Hirzebruch surfaces $\IF_{n_r^i}$ blown up along the
intersection of $t^i_r$ with a fiber (fiber over the same point of $\IP^1$
for all $i$).  The proper transforms are denoted by $\widetilde{t^i_r}$.
There are also the exceptional curves $u^i_r$
introduced by the blowups.

$E^i_r$ and $E_r^{i+1}$ are glued together
by attaching the sections $\widetilde{t^i_r}$ to $s_r^{i+1}$.
Since the normal bundles of this common curve in $E^i_r$ and $E_r^{i+1}$
can be seen to add up to~$-2$, this can be embedded locally
in a Calabi-Yau threefold.

Finally $E^i_r$ meets $E_{r+1}^i$ by the identification of
$u_r^i$ with $u_{r+1}^i$; these are each curves with
self-intersection $-1$, so the
union can be locally embedded in a Calabi-Yau threefold.
Note that we have now one fiber with $k+M-1$ components.
The Mori cone is generated by these $k+M-1$ components of
the fiber, together with $M$ sections $s_r$, giving
the total of $k+2M-1$ generators.

\newsec{Trivalent Geometry and Addition of Fundamental Matter}

To add $N_f$ matter in the fundamental representation for a given
gauge group $SU(N_c)$, we can introduce an extra sphere which intersects
a given ${\bf P}^1$ base corresponding to the gauge group of interest,
on top of which we have an $A_{N_f-1}$ singularity.  In this
way we obtain extra bi-fundamental matter $(N_f,N_c)$.  By degenerating
the extra sphere, i.e. by considering the extra sphere to be very large,
we weaken the $SU(N_f)$ dynamics, thus `demoting' it to a spectator
flavor symmetry group.   In this section we would like
to add matter to the linear chain of $SU$ groups considered
in the previous section.
 In order to construct this geometry along the lines just
discussed
we clearly need an additional building
block of our base geometry, corresponding to the trivalent
vertices in the base. That is we need a central
sphere of self-intersection $-2$ intersecting three other
2-spheres once.

A non-trivial constraint arises from the fact that this
geometry has to be compatible with the Calabi--Yau
condition. Recall that each Mori vector defines a 2-cycle $C_i$
which is a generator of the canonical base for the 2-cycle homology
and moreover that the intersections of this 2-cycle with
the divisors in $X$ are proportional to the entries of the
Mori generator. If $X$ is Calabi--Yau, the entries of each Mori
generator always have to add up to zero\foot{The fact that
$c_1=0$ implies $\sum_i \lm{a}{i}=0$ is a basic result in
toric geometry. Alternatively, in the linear sigma model
language, $\sum_i \lm{a}{i}$ describes the anomaly
contribution to the $a$-th $U(1)$ factor \qsum.}. The Calabi--Yau
geometry forces us therefore to introduce one other vertex
with contribution $-1$ such that the Mori vector of the central
2-sphere becomes $\lm{0}{}=(-2,1,1,1,-1,0,\dots)$. The local mirror
geometry, solving \lmm, contains now the monomials
\eqn\mdn{
1,x,y,z,xyz \ ,
}
where we have used a $\ICs$ action of the solution to scale one of
the variables to 1.

The solution to the local mirror geometry of the trivalent vertex
uses two more variables than the 2-vertex we used previously
in the chain. If we want to keep the
dimension of the manifold $X$ three, adding trivalent vertices,
we have to add more equations, that is, we have to consider a
complete intersection manifold.

\subsec{Gauge Group $SU(N)$ with $M$ Fundamental
Matter Revisited}
Before proceeding to the case of linear chain with extra
fundamental matter, let us consider the special case of
$SU(N)$ with $M$ fundamental matter.  We have
already done this by viewing it as a special case of
the chain of $SU(N)\times SU(M)$ where the $SU(M)$ coupling
is weakened.  Now we wish to obtain it again using
the trivalent geometry constructed above. The type IIA
geometry is shown in fig. 9a).
\vskip 0.5cm
{\baselineskip 12pt\sl
\goodbreak\midinsert
\centerline{\epsfxsize 4.2truein\epsfbox{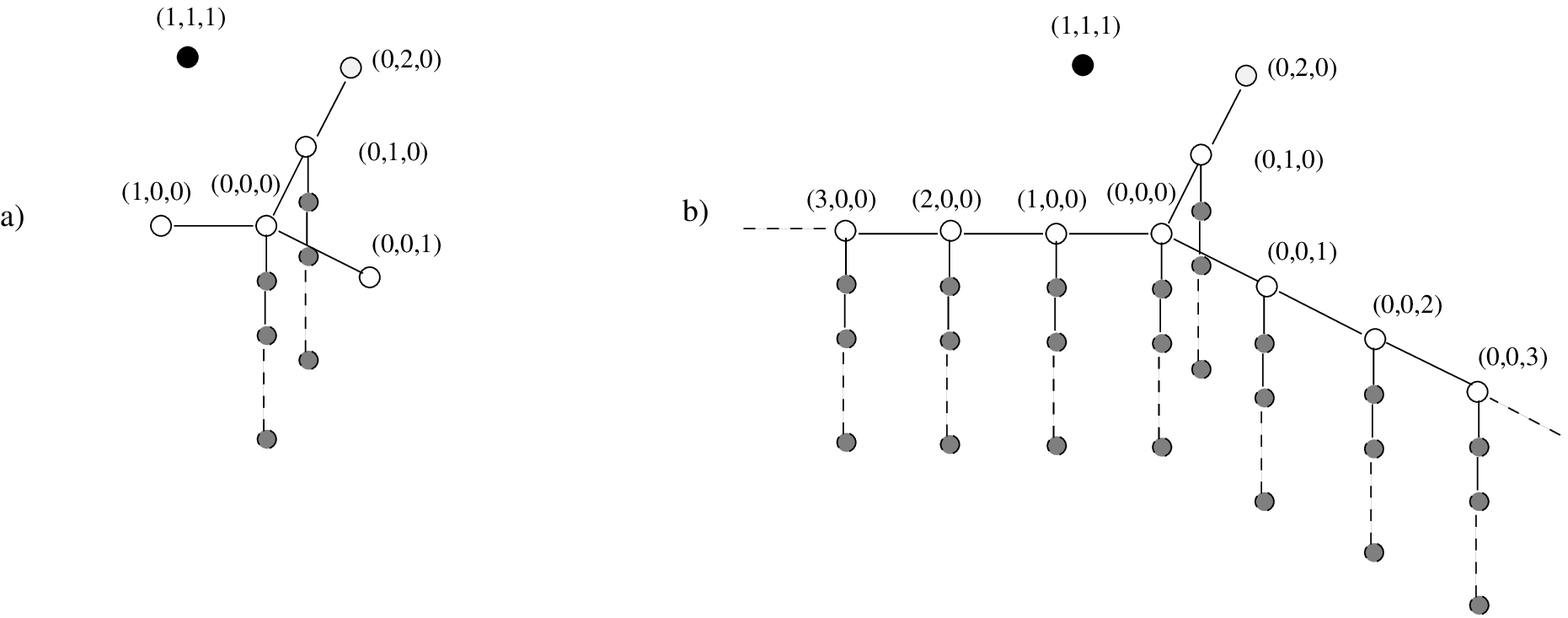}}\leftskip 1pc
\rightskip 1pc\vskip0.3cm
\noindent{\ninepoint  \baselineskip=8pt {\bf Fig. 9:}
a) $SU$ gauge group with fundamental matter from the trivalent
vertex geometry,\
b) Product gauge group $\prod SU(l_r)\ \cdot \
SU(n)\ \cdot \
\prod SU(k_r)$ with bi-fundamentals and fundamental matter
in the $SU(n)$ factor.}
\endinsert}\ni
The vertices for the base geometry can be read off from the figure
(we add an extra zero at the end of each vertex for the fiber
direction).
The fiber geometry involves an $A_{N-1}$ chain of $N-1$ 2-spheres fibered
over $s_{0,0,0}$, the central base $\IP^1$, generating the gauge symmetry,
and a further $A_{M-1}$ chain of $M-1$ 2-spheres fibered over another base
$\IP^1$, $s_{0,1,0}$ (in an obvious notation), which generates
the matter. These two chains are described by two sets of vertices
$\ninepoint \nu^{0,0,0}_k=(0,0,0,k),\ k=1,\dots,N$ and
$\ninepoint \nu^{0,1,0}_k=(0,1,0,k),\ k=1,\dots,M$, respectively.
By sending the gauge coupling of the $SU(M)$
gauge group to zero, the ``degeneration'' process described
previously, we are left only with $SU(N)$ gauge group
with $M$ fundamentals.
Geometrically we take the base $\IP^1$ of the $A_{M-1}$ factor to have
infinite size\foot{Note that the 2-cycles supporting the matter
multiplets sit at a common point of this $\IP^1$. Intuitively
the limit corresponds to restricting to the neighborhood of
this point on the base $\IP^1$ while deleting the dependence on the
global structure - this will effectively lead to a reduction of the dimension
of $X$.}.
In the toric geometry, this is easily done by simply cutting
the length of the leg carrying the $A_{M-1}$ chain to one, that is deleting
the vertex $\nu^{0,2,0}$ from our polyhedron.

We proceed with the solution of the model. The $(N)+(M)+(4+1)$
vertices fulfill $N+M$ relations corresponding to the
$N-1$ Coulomb parameter $\zc_i$ of $SU(N)$,
$M$ mass parameters $\zm_i$ and the universal scale parameter $\zg$:

\eqn\caseiirel{
\ninepoint
\zc_i = \ov{a^{0,0,0}_{i-1}a^{0,0,0}_{i+1}}{(a^{0,0,0}_{i})^2},
\
\zm_0 = \ov{a^{0,0,0}_{0}a^{0,1,0}_{1}}{a^{0,0,0}_{1}a^{0,1,0}_{0}},
\
\zm_i = \ov{a^{0,1,0}_{i-1}a^{0,1,0}_{i+1}}{(a^{0,1,0}_{i})^2},
\
\zg = \ov{a^{0,1,0}_0a^{1,0,0}_0a^{0,0,1}_0}{(a^{0,0,0}_0)^2a^{1,1,1}_0}.
}
The mirror geometry $\xs$
is described by the hypersurface
\eqn\caseiipol{\eqalign{
&p(\xs)=P_{N}(w)+x+y+ z\;(a^{1,1,1}_0\; xy+ Q_{M}(w)), \cr
&P_{N}(w)=\sum_{i=0}^{N} a^{0,0,0}_i\; w^i,\qquad
Q_{M}(w)=\sum_{i=0}^{M} a^{0,1,0}_i\; w^i \ ,
}}
where we have set various coefficients to
one using the $\ICs$ symmetries of the solution.
Note that $z$ appears only linearly in $p(\xs)$ and can be integrated out.
This integration imposes the constraint
$$
e=xy+Q_{M}(w)=0 \ ,
$$
and solving $e=0$ for, say,  $y$, we get finally
$$
p(\xs)=x+P_{N}(w)+\ov{Q_{M}(w)}{x}  \ ,
$$
which is the expected result.

\subsec{Product Gauge Group $\prod SU(k_r)$
with $(k_r,\bar{k}_{r+1})$ $\oplus \; m_r\cdot (k_r)$ matter}
The above construction generalizes straightforwardly to the
product gauge groups considered previously.
For each single group factor $SU(k_r)$ we replace the
base $\IP^1$ by the central $\IP^1$ of the trivalent
geometry, as shown in fig. 9b) for one $SU$ factor.

The base geometry is now a chain of $T$ trivalent vertices,
where above each central $\IP^1$ we have an $A_{k_r-1}$ chain
of 2-spheres generating the gauge symmetry and above each
layer $\IP^1$ there is an $A_{m_r-1}$ chain of spheres adding
the fundamental matter.
The vertices of the toric type IIA geometry can be obtained from
plumbing together those of a single trivalent vertex as described
before. They can be also recovered from our description of the
type IIB mirror geometry below.
\vskip 0.5cm
{\baselineskip 12pt\sl
\goodbreak\midinsert
\centerline{\epsfxsize 2.6truein\epsfbox{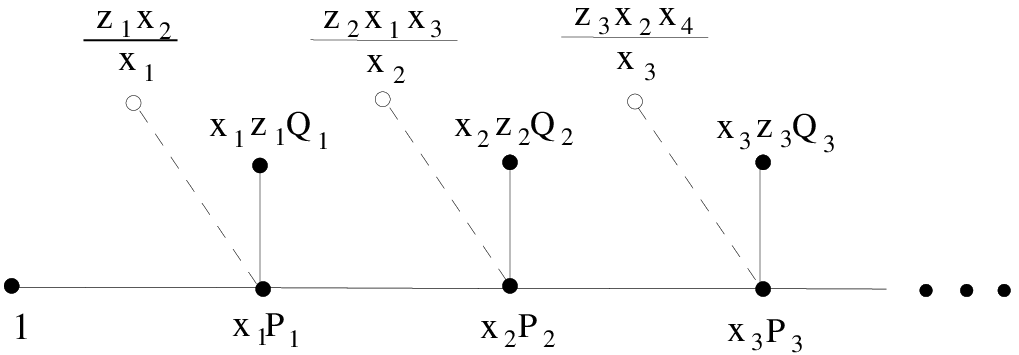}}\leftskip 1pc
\rightskip 1pc\noindent{\ninepoint  \baselineskip=8pt {\bf Fig. 10}
}
\endinsert}\ni

Fig. 10 shows the structure of our chain of trivalent vertices,
where we have indicated the monomials that solve the mirror
geometry. The monomials associated to the geometry of the $r$-th
trivalent vertex can be collected into three groups:
$$
\vbox{\offinterlineskip\tabskip=0pt\halign{\strut
$#$~~~\ \hfil &$#$~\hfil \cr
x_rP_r(w),&P_r(w)=\sum_{i=0}^{k_r}\; p^r_i\; w^i\ ,\cr
x_rz_rQ_r(w),&Q_r(w)=\sum_{i=0}^{m_r}\; q^r_i\; w^i\ ,\cr
\rho_r=\mu_r\; x_{r-1}x_{r+1}\ov{z_r}{x_r}\ ,&\cr}}
$$
where $x_{T+1}$ and $x_r,z_r,\ r=1,\dots,T$ are $2\; T+1$ coordinates
for the base geometry and $w$ is the fiber variable. The
polynomials
$P_r(w)$ and $Q_r(w)$ describe the $r$-th gauge system and its
associated fundamental matter, respectively.
Moreover
$p^r_i,\; q_i^r$ and $\mu_r$ are parameters related to the moduli
of the combined gauge and matter system. The monomials
combine to the hypersurface constraint
\eqn\caseiiipol{
p(\xs)=1+\sum_{r=1}^{T+1} x_r P_r(w)+\sum_{r=1}^T x_rz_rQ_r(w)
+ \sum_{r=1}^T \rho_r \ .
}
The coordinates $z_r$ appear all linearly and integrating them out
imposes a set of $T$ constraint equations:
$$
e_r:\ \ x_r^2Q_r(w)+\mu_rx_{r-1}x_{r+1}=0 \ ,r=1,\dots,T,
$$
with solution
\def\tQ{\tilde{Q}}
$$
x_r=x^r\; \prod_{\al=1}^r
(\tQ_\al)^{r-\al}(w),\qquad \tQ_r(w)=-\ov{1}{\mu_r}Q_r(w) \ .
$$
Plugging them back into \caseiiipol\ we obtain the final
mirror geometry, which describes again a Riemann surface:
\eqn\caseiiipolii{
p(\xs)=1+\sum_{r=1}^{T+1} x^rP_r(w)\prod_{\al=1}^r (\tQ_\al)^{r-\al}(w) \ .
}
\subsubsec{Exact Solution for the Coulomb Branch}
The Riemann surface \caseiiipolii\ has the geometrical interpretation of a
single type IIA five-brane \klmvw\
as a result from a T-duality transformation on the ALE fiber space \oov.
For the present gauge and matter content it was derived
in \witex\ from strong-weak coupling duality between type IIA and
M-theory. However the exact solution of the gauge theory needs
more than just the geometry of the Riemann surface $E$,
as is clear from the fact that the same $E$ can describe
very different kind of $N=2$ theories depending on a choice
of metric on $E$. These data can be obtained easily from the
type IIB picture, which was the starting point of the T-duality
to the five-brane in \klmvw.  This is a straight-forward aspect of our toric
construction:
once the type IIA geometry is set up, the toric machinery proceeds
in a stubborn and precise way until the very end, giving the exact solution
of the theory in terms of period integrals on $\xs$.

The moduli of the above gauge/matter system in terms of
complex structure moduli of type IIB are given by
$$\ninepoint\eqalign{
\zc_{r,i}&=\ov{p^r_{i-1}p^r_{i+1}}{(p^r_i)^2},
\ i=1,\dots,k_r-1\cr
z^{(m),fund}_{r}&=\ov{p^r_{0}q^r_{1}}{p^r_{1}q^r_{0}},\ 
z^{(m),fund}_{r,i}=\ov{q^r_{i-1}q^r_{i+1}}{(q^r_i)^2},\
\ i=1,\dots,m_r-1\cr
z^{(m),bifund}_r&=\ov{p^r_0p^{r+1}_1}{p^r_1p^{r+1}_0},\cr
\zg_r&=\ov{p_0^{r-1}p_0^{r+1}q_0^{r}}{(p_0^{r})^2\mu_r}}
$$
They describe the complex volumes of 2-spheres as follows:
$\zc_{r,i}$ are the $k=\sum k_r-1$ Coulomb parameters,
$z^{(m),fund}_{r,i}$ and $z^{(m),fund}_{r}$ describe the $m=\sum m_r$
mass parameters for the fundamental matter and
$z^{(m),bifund}_r$ describe the $T$ mass parameters for the bi-fundamentals.
Moreover $\zg_r$ are $T$ coupling constants, parameterizing the
$T-1$ relative gauge couplings and the Planck mass.

For the derivation of the periods from the Picard-Fuchs
equations, it is convenient to use a 
complete intersection description of the above geometry 
\ref\bor{V. Batyrev and L. A. Borisov, {\it On Calabi--Yau
complete intersections in toric varieties}, alg-geom/9412017
}.
To do this, we have to give the so-called
nef partition, which roughly speaking corresponds to finding
subsets $S_r$ of fields whose $U(1)$ charges $q_i^a$ 
add up to zero within each set. For the present case this is trivial 
due to the
linear appearance of the $z_r$ in \caseiiipol, 
and the nef partition is obtained by
grouping the vertices and monomials by powers of the $z_r$. In this
way we obtain a set $S_0$ containing all $z_r$ independent 
monomials and $T$ sets $S_r$ containing the monomials linear
in $z_r$ for each $r$. The holomorphic $(3,0)$ form is then
\ref\batx{V.V. Batyrev, Duke Math.\ J.\ {\bf 69} (1993) 349}%
\hkty
$$
\Omega = \ov{1}
{P_0} \prod_{r=1}^T \ov{1}{P_r }\ 
\prod_{i=1}^{T+4}\ov{dX_m}{X_m},
$$
where the $T+4$ variables $X_m$ correspond to the base 
variables $x_r,r=1,\dots,T+1$, the fiber
variable $w$, and two extra trivial variables $v,u$ which we add 
quadratically to $P_0$ as in as in \lch.
Here $P_0$ is the Laurent polynomial related to the hypersurface
constraint by 
setting $z_r=0,\forall r$ in \caseiiipol\ and 
similarly the $P_r$ correspond to the $T$ equations $e_r$\foot{For
the precise definition of Laurent polynomials and the coordinates
$X_m$, see \batx. For an alternative canonical formulation using 
homogeneous coordinates, see \ref\batcox{ V.\ Batyrev and D.\ Cox,
Duke Math.\ J.\ {\bf 75} (1994) 293}.}. 
More precisely, the residue of $\Omega$ gives the holomorphic
$(3,0)$ form on the threefold defined by $P_r=0,\ r=0,\ldots,T$.

It is straightforward to check that $\Omega$ fulfills 
the GKZ system of differential equations \gkz, 
which determine the instanton expansion for 
the period integrals of $\Omega$.

Let us finally give the field theory limit and check that it
agrees with the general expectations. Requiring to be near
the singular point of each of the $A$ factors of the fiber
geometry, the limit is of the form
$$
p^r_0 \sim \ep^{-k_r},\ q^r_0 \sim \ep^{-m_r},\ \mu_r \sim \ep^0 \ ,
$$
implying finiteness of the field theory moduli
$$
\zc_{r,i} \sim z^{(m),fund}_{r,i} \sim z^{(m),fund}_{r}
\sim z^{(m),bifund}_r \sim \ep^0 \ .
$$
On the other hand the scale variables $\zg_r$ behave as
$$
\zg_r \sim \ep^{-b_r},\qquad b_r=2k_r-k_{r+1}-k_{r-1}-m_r \ ,
$$
precisely as required by the beta-function coefficient $b_r$ of the
$r$-th gauge factor.

\newsec{Affine $E_n$ Geometries From the Trivalent Geometry}
In the next section we will derive the mirror geometry of
affine singularities using elliptic fibrations over
the complex plane.
Here we want to describe briefly, how we can get the affine
$E_n$ geometries from the trivalent geometry by extending
what was done in the previous section.

Since we are interested in the base geometry to look
like an affine $E_n$ Dynkin diagram, we now allow
all three $A$ chains emerging
from the central sphere have length $>1$. Let us denote
these chains as $A_{p-1},\ A_{q-1},\ A_{r-1}$. From \mdn\ and the obvious
intersections of the $A$ chains, the mirror geometry
contains the monomials
\eqn\monpqr{
1,x,y,z,xyz;\ x^2,x^3,\dots,x^p;\ y^2,y^3,\dots,y^q;\ z^2,z^3,\dots,z^r.
}
and describes the complex deformation of a
singularity called $T_{p,q,r}$
\ref\arn{See e.g., V.\ Arnold, A.\ Gusein-Zade and A.\ Varchenko,
{\it Singularities of Differentiable Maps I, II}, Birkh\"auser 1985.}.
{}From the discussion of the trivalent vertex in the
previous section, it is clear what this corresponds
to the mirror of type IIA geometry of trivalent base geometry.
\vskip 0.5cm
{\baselineskip=12pt \sl
\goodbreak\midinsert
\centerline{\epsfxsize 1.8truein\epsfbox{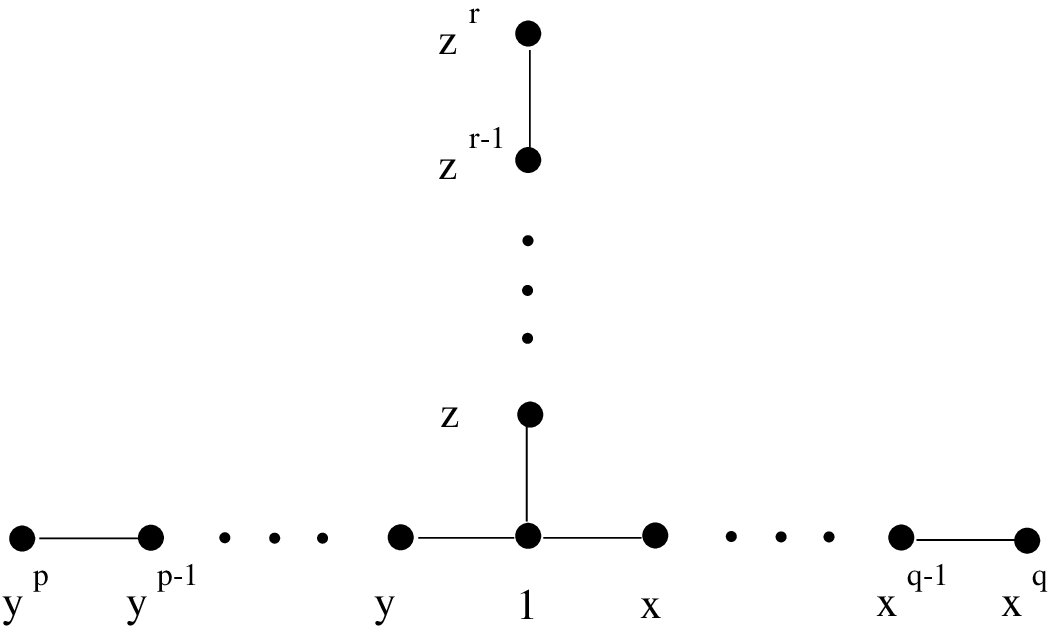}}
\leftskip 1pc\rightskip 1pc\vskip0.3cm
\noindent{\ninepoint{\bf Fig. 11:}
The $T_{p,q,r}$ singularity.}
\endinsert}\ni
The general singularity of this type
has indefinite intersection form and will not lead to the
perturbative prepotential of a well-defined field theory in four dimensions
(though it would be of interest for questions of mirror symmetry
in $N=4$ theories in $d=3$ as discussed later in the paper).  As
discussed in section 2 the limiting case that would be of interest
for constructing superconformal theories in
four dimensions are the ones (with semidefinite intersections)
\eqn\jockoalt{\eqalign{
T_{3,3,3}&:y^3+x^3+z^3+\mu xyz,\cr
T_{2,4,4}&:y^2+x^4+z^4+\mu xyz,\cr
T_{2,3,6}&:y^2+x^3+z^6+\mu xyz \ ,
}}
which give affine $E_6,E_7$ and $E_8$ geometry in the base.
We can now proceed to the construction of the $E_n$ type of
superconformal theories as follows: we start with a type IIA
singularity which is composed of three $A_{n-1}$ chains of
length $n_i=(p,q,r)$, which intersect the central sphere
of a trivalent geometry, with $(p,q,r)$ being one of the values
in \jockoalt. For concreteness let us consider the $E_6$ case corresponding
to $(p,q,r)=(3,3,3)$. The mirror geometry is described by the monomials
\monpqr, combined in the hypersurface constraint
\eqn\affies{
p(\xs)=v^3+v^2(x+y+z)+v(x^2+y^2+z^2)+x^3+y^3+z^3 +xyz\ ,
}
where $(x,y,z,v)$ are the homogeneous variables.
Eq. \affies\ describes a del Pezzo surface $B_6$ with $c_1\neq0$.
We can easily restore $c_1=0$ by using $v\cdot p(\xs)$ as
the hypersurface constraint, which describes a singular quartic
$K3$ surface.

However this is not the end of the
story because we wish to use this $B_6$ geometry
as the base and fibering $A_n$ chains above the
blow up spheres of the original type IIA geometry. Let $k_{x,i}$
denote the rank of the $A_n$ chain above the $i$-th base sphere in the $x$
direction and similarly for $y,z$. The mirror geometry becomes
\eqn\affiesb{\eqalign{
p(\xs)=&v^3P_{k_{x,0}}+v^2(xP_{k_{x,1}}+yP_{k_{y,1}}+zP_{k_{z,1}})+
\cr&v(x^2P_{k_{x,2}}+y^2P_{k_{y,2}}+z^2P_{k_{z,2}})+x^3+y^3+z^3 +xyz\ ,
}}
where the coefficients $P_i=P_i(w,w^\prime)$ are polynomials
of degree $i$ in the homogeneous variables $w,w^\prime$
on the fiber. For
$$
k_{x,i}=k_{y,i}=k_{z,i}=(3-i)k,\ k\in \IZ,
$$
the hypersurface determined by the polynomial $vww^\prime p(\xs)=0$
describes a
Calabi--Yau threefold $\xs$. There are
two independent $\ICs$ actions $(y,x,z,v,w,w^\prime)$
$\to$
$(\lambda^{k-2}y,\lambda^{k-2}x,\lambda^{k-2}z,$
$\lambda^{-2}\mu^{-k}v,\lambda\mu w,\lambda\mu w^\prime)$,
which can be used to set $v$ and $w'$ to one.

The above geometry, and its $E_7,\ E_8$ variants based on the other
two singularities in \jockoalt,
describe the exact solutions to
the superconformal $N=2$ gauge theories defined in sect. 2.
The toric geometry of the $E_n$ base can be read off from fig. 12,
where to each node we show the associated monomial of the mirror geometry
and the Dynkin numbers which determine the relative multiplicities
in the rank of the $A_n$ fibers.

\vskip 0.5cm
{\baselineskip=12pt \sl
\goodbreak\midinsert
\centerline{\epsfxsize 4.6truein\epsfbox{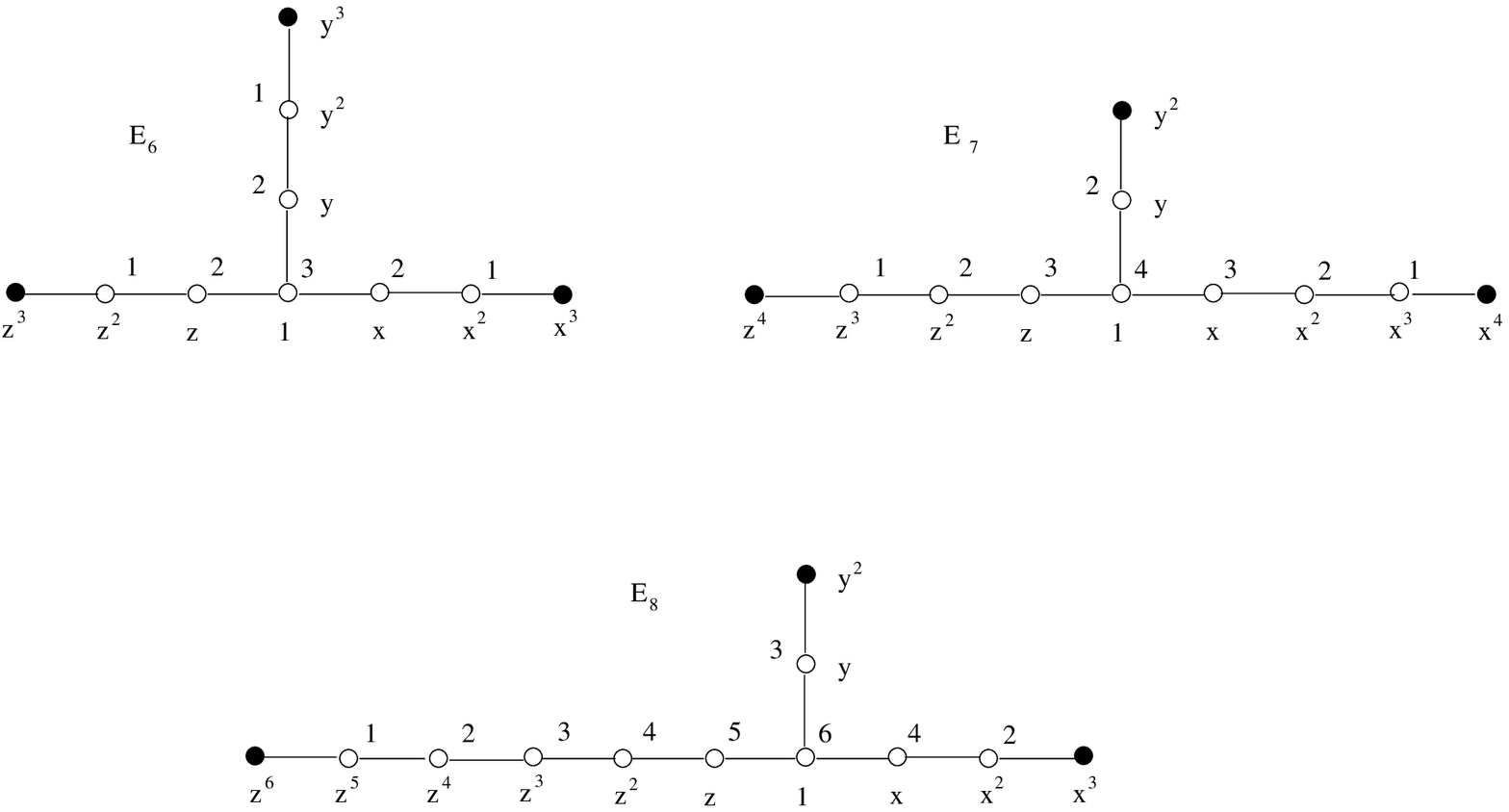}}
\leftskip 1pc\rightskip 1pc\vskip0.3cm
\noindent{\ninepoint{\bf Fig. 12:}
Affine $E_n$ base geometry of the 4d superconformal theories.}
\endinsert}\ni
The fibration of the $A_n$ fibers proceeds as in the $E_6$ case
described above and leads to the following defining polynomials
for the mirror geometry:
\eqn\affenallpol{{\ninepoint\eqalign{
&\hskip -1cm E_6:\ SU(k)^3 \times SU(2k)^3 \times SU(3k)\cr
&\hskip -1cm p(\xs)=y^3+x^3+z^3+\mu xyz\cr&\ \ +\
\sum_{i=1}^3z^{3-i}v^iP^z_{i\cdot k}(w)
+\sum_{i=1}^2x^{3-i}v^{i}P^x_{i\cdot k}(w)
+\sum_{i=1}^2y^{3-i}v^{i}P^y_{i\cdot k}(w),\cr
&\hskip -1cm E_7:\ SU(k)^2 \times SU(2k)^3 \times SU(3k)^2 \times SU(4k)\cr
&\hskip -1cm p(\xs)=y^2+x^4+z^4+\mu xyz\cr&\ \ +\
\sum_{i=1}^4z^{4-i}v^iP^{z}_{i\cdot k}(w)
+\sum_{i=1}^3x^{4-i}v^{i}P^x_{i\cdot k}(w)
+\sum_{i=1}^1y^{2-i}v^{2i}P^y_{2i\cdot k}(w),\cr
&\hskip -1cm E_8:\ SU(k) \times SU(2k)^2 \times SU(3k)^2
\times SU(4k)^2 \times SU(5k)\times SU(6k) \cr
&\hskip -1cm p(\xs)=y^2+x^3+z^6+\mu xyz\cr&\ \ +\
\sum_{i=1}^6z^{6-i}v^iP^{z}_{i\cdot k}(w)
+\sum_{i=1}^2x^{3-i}v^{2i}P^x_{2i\cdot k}(w)
+\sum_{i=1}^1y^{2-i}v^{3i}P^y_{3i\cdot k}(w).}}}

\subsec{Exact Solution for the Coulomb Branch}
We proceed with the exact solution of the superconformal
theory corresponding to the affine $E_8$ quiver. The other
two cases can be treated very similarly.

To fix notations, let us label the vertices of the type IIA
geometry in fig.12 by the letters $a_{i,j}$ for the ``$z$'' leg
and similarly by $b_{i,j}$ for the other two legs starting from the
central sphere.
The first subscript $i$ denotes the Dynkin number and the second
subscript the $j$-th vertex of the $A_n$ chain of the fiber.
The fiber polynomials in \affenallpol\ are then
$$\eqalign{
P^x_{2i\cdot k}(w)&=\sum_{j=0}^{2i\cdot k} b_{i,j}w^j,\ i\in\{2,4\},\quad
P^y_{3k}(w)=\sum_{j=0}^{3k} b_{3,j}w^j,\cr
P^z_{i\cdot k}(w)&=\sum_{j=0}^{i\cdot k} a_{i,j}w^j,\ i={1,\dots,6}.
}$$
We abbreviate the parameters of the base by dropping the second
subscript, $a_i\equiv a_{i,0}$, etc.
In total we have $13+c_2(E_8)k=30k+13$ vertices. To obtain a local
threefold they should fulfill $30k+8$ relations.
These are the $30k-9$ Coulomb parameters, 8 mass parameters,
the 8 relative coupling constants and the elliptic modulus:
\def\dots{...}
\eqn\mode{{\ninepoint\eqalign{
\zg_i&=\ov{a_{i+1}a_{i-1}}{a_i^2},i=1,\dots,5,\ a_0\equiv 1,\quad
\zg_6=\ov{a_5b_3b_4}{a_6^2\mu},\cr
\yg_2&=\ov{b_4}{b_2^2},\ \yg_3=\ov{a_6}{b_3^2},\ \yg_4=\ov{a_6b_2}{b_4^2},
\cr
\zc_i&=\ov{a_{i,l-1}a_{i,l+1}}{a_{i,l}^2},i=1,\dots,6,\ l=1,\dots,ik-1,
\cr
\yc_i&=\ov{b_{i,l-1}b_{i,l+1}}{b_{i,l}^2},i=2,3,4,\ l=1,\dots,ik-1,
\cr
\zm_i&=\ov{a_{i,0}a_{i+1,1}}{a_{i,1}a_{i+1,0}},i=1,\dots,5,\
\zm_6=\ov{b_{4,0}b_{2,1}}{b_{4,1}b_{2,0}},\
\zm_7=\ov{a_{6,0}b_{4,1}}{a_{6,1}b_{4,0}},\
\zm_8=\ov{a_{6,0}b_{3,1}}{a_{6,1}b_{3,0}} \ ,
}}}
where we have used a basis adapted to the Dynkin diagram for the
coupling constants $\zc_i$. The elliptic class is given by
$\prod_{k=1}^6(\zg_k)^k\prod_{k=2,3,4}(\yg_k)^k=\mu^{-6}$.

Using the definition \algc\ of the complex structure moduli,
it is easy to read off the charge vectors $\lm{a}{}$ from \mode.
The exact instanton corrected prepotential is then given in
terms of the period integrals, the solution to the Picard-Fuchs
system \gkz. The field theory limit is as in eq.\ftl, that is
the coefficient of the power of $w^i$ of a $SU(k)$ factor 
scales as $\ep^{k-i}$. It is easy to check that all moduli
in \mode\ scale as $\ep^0$ in this limit. Moreover the
holomorphic form on $X$ can be written in homogeneous 
variables as
$$
\Omega = \ov{dydxdzdvdwdw'}{vww'p(\xs)},
$$
where $w'$ is the second homogeneous variable on the
fiber. More precisely, taking into account the invariance under the
two $\ICs$ actions, $\Omega$ restricts to the holomorphic $(3,0)$ form 
on the hypersurface $p(\xs)=0$.
Very similar statements hold for the field theory limits
and the holomorphic forms for the following theories based on 
affine base geometries and will not be repeated.

It is clear from the above that we can also obtain
the curve for the case with arbitrary ranks of the gauge
group corresponding to the affine $E$ base, simply by changing
the degree of the corresponding functions of $w$.
We can also obtain the exact solution if we put additional
fundamental matter for each gauge group, as we will next discuss.

\subsec{$D$ and $E$ Dynkin Diagrams as the Base}
As noted in section 2, asymptotic freedom also
allows $SU$ gauge theories based on ordinary $D$ and $E$ Dynkin
diagrams.
Note that the ordinary
$D$ and $E$ Dynkin diagrams also correspond
to $T_{p,q,r}$ geometry. In particular $D_n$ corresponds
to the $T_{2,2,n-2}$ geometry and the ordinary $E$ cases
are the same as $T_{2,3,5}$,$T_{2,3,4}$ and $T_{2,3,3}$
for $E_{8,7,6}$ respectively.  It is clear from the previous
discussion how one writes the type IIB geometry for these
cases with arbitrary rank gauge group on top of each node.
Moreover, we can add fundamental matter to each gauge group just
as in the case of linear chain.  In fact treating
each edge of the $E_n$ or $D_n$ Dynkin diagram emanating from the
trivalent vertex as a linear chain, we have a situation
already studied.  In solving for the curve
we will get just as in the linear chain case additional
polynomials raised to some powers
for each node, and their powers will increase as we go down
the chain.
Since various aspects of these have already been discussed in detail
in previous sections we will not go into
any further detail.

\newsec{Elliptic Singularities and Affine ADE Quivers}
We complete now the description of gauge groups
with the base geometry given by affine ADE singularities
using elliptic fibrations over the complex
plane. This kind of singularities is well studied
in the mathematics literature \kod\mir\
and has been analyzed thoroughly in the context of
heterotic/F-theory duality in \sixau.

The result from local mirror symmetry we obtain is that
{\sl the local mirror geometry of the K\"ahler resolution of
affine ADE singularity describes the moduli space of
flat ADE bundles on a torus $E$.} This was already anticipated
from our discussion of section 3.
 Although we restrict again
to the ADE case, the non-simply laced cases
can be treated similarly \pape.

We proceed in this section as follows. First we will study
the missing cases,
the two infinite series of 4d superconformal theories
based on the affine $A_n$ and $D_n$ geometries, respectively.
We will then relate our results for all the $ADE$
geometries to the description of flat $G$
bundles on elliptic curve $E$  and use this connection to
determine the $S$ duality groups.

\subsec{Affine $A_n$ from Elliptic Fibrations over the Plane}
The first of the two infinite series of superconformal 4d gauge theories
consists of fibering $A_k$ singularities over the affine $A_N$ base.
This base geometry can be constructed by blowing up the
local singularity
\eqn\afansing{
y^2+x^3+x^2+t^{N+1}
}
in a fibration of an elliptic curve $E$ over the $\IC$ plane.
As before, we choose a sextic in $\WP_{1,2,3}$ as our model
of $E$.

Let us describe the toric data for the type IIA base geometry.
The polyhedron is defined by four vertices $\vt$, describing
the generic sextic, and $N+1$ vertices $v_i,\ i=1,\dots,N+1$ introduced by
the blow ups:
\eqn\afanvert{{\ninepoint\eqalign{
&\vt_0=(0,0,0),
\vt_1=(0,2,3),
\vt_2=(0,-1,0),
\vt_3=(0,0,-1),\cr
&v_1=(1,2,3),v_{2i}=(1,2-i,3-i),v_{2i+1}=(1,2-i,2-i),\ i>0}}}

Above each blow up sphere we fiber now an $A_k$ singularity.
To describe the toric polyhedron, we add a zero entry at the
end of each vertex in \afanvert\ and join
$(k+1)\cdot (N+1)$ further vertices:
\eqn\afanvertii{{\ninepoint\eqalign{
v_{i,l}=v_i+(0,0,0,l),\ l=1,\dots,k+1,}}}
This completes the construction of the type IIA geometry.

\vskip 0.5cm
{\baselineskip=12pt \sl
\goodbreak\midinsert
\centerline{\epsfxsize 1.7truein\epsfbox{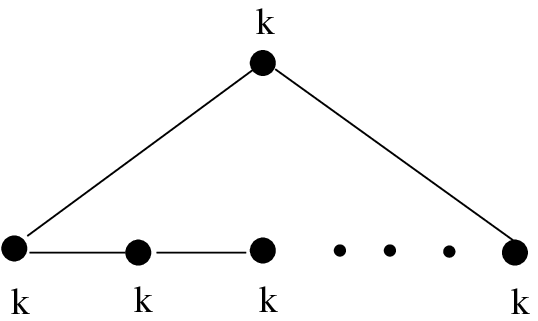}}
\leftskip 1pc\rightskip 1pc\vskip0.4cm
\noindent{\ninepoint  \baselineskip=8pt {\bf Fig. 13:}
$SU(k+1)^{N+1}$ gauge theory from affine $A_N$ in the base.
}\endinsert}\ni
As before, let us associate to each vertex $v_{i,l}$
the parameter $a_{i,l}$ that multiplies the  corresponding
monomial in the defining equation of the mirror geometry.
Moreover we abbreviate $a_{i,0}$ by $a_i$.
The relations $\lm{a}{}$ needed to define the mirror geometry
and its moduli space, written in terms of the moduli \algc,
are as follows: The volumes in the base are parametrized by
the relations:
\eqn\afanmod{{\ninepoint\eqalign{
\zg_1&=\ov{a_2a_3\at_1}{a_1^2\at_0},\quad
\zg_2=\ov{a_1a_4}{a_2^2},\quad
\zg_3=\ov{a_1a_5\at_3}{a_3^2\at_0},\cr
\zg_{2i+2}&=\ov{a_{2i}a_{2i+4}}{a_{2i+2}^2},\ i=1,\dots,
\cases{\ov{N}{2}-2& $N$ even\cr \ov{N+1}{2}-2& $N$ odd},\cr
\zg_{2i+1}&=\ov{a_{2i-1}a_{2i+3}}{a_{2i+1}^2},\ i=2,\dots,
\cases{\ov{N}{2}-1& $N$ even\cr \ov{N+1}{2}-2& $N$ odd},\cr
\zg_N&=\ov{a_{N-2}a_{N+1}}{a_N^2}\times
\cases{\ov{\at_2}{\at_0}&$N$ even\cr \ov{\at_3}{\at_0}&$N$ odd},\quad
\zg_{N+1}=\ov{a_{N}a_{N-1}}{a_{N+1}^2}\times
\cases{\ov{\at_2\at_3^2}{\at^3_0}&$N$ even\cr
\ov{\at_3\at_2^2}{\at^3_0}&$N$ odd}
}}}
Moreover, the gauge system of the fiber is described by
$k\cdot (N+1)$ Coulomb moduli and $N+1$ mass parameters,
of which $N$ are independent:
\eqn\afanmodii{
\zc_{i,l}=\ov{a_{i,l-1}a_{i,l+1}}{a_{i,l}^2},\qquad
\zm_{i,l}=\ov{a_{i,0}a_{i+2,1}}{a_{i,1}a_{i+2,0}}.
}
In total we have $n_\nu=4+(N+1)\cdot(k+2)$ vertices fulfilling
$n_R=1+N+(N+1)\cdot k +N$ relations. Taking into account
the hypersurface constraint this gives a mirror geometry of
dimension $n_\nu-n_R-2=3$.
Combining the monomials that solve \lmm\ with the $\lm{a}{}$ as
defined by the above relations and \algc, the hypersurface constraint
reads:
\eqn\affanpol{\eqalign{
A_N\ :\ &SU(k+1)^{N+1}\cr
\ p(\xs) &= v(y^2+x^3+z^6+\mu yxz)\cr&
+z^{N+1} P^{(N+1)}_{k+1}(w)+z^{N-1}x P^{(N-1)}_{k+1}(w)+z^{N-2}y P^{(N-2)}_{k+1}(w)
+\dots\cr&+\cases{
yx^{\ov{N-2}{2}}P^{(0)}_{k+1}(w)& $N$ even\cr
x^{(N+1)/2} P^{(0)}_{k+1}(w)& $N$ odd\cr}
}}

\ni
The coefficients
in the polynomials $P^{(K)}_{k+1}(w)$ are related to those 
in eqs.\afanmod,\afanmodii\ by 
$P^{(K)}_{k+1}(w)=\sum_{l=0}^{k+1}a_{N+1-K,l}\; w^l$.

To solve $p(\xs)$, note that $v$ appears only linearly and can be integrated
out. This results in the constraint:
$$
E:\ y^2+x^3+z^6+\mu yxz=0 \ .
$$
Thus $(y,x,z)$ become coordinates on an elliptic curve $E$. We are left
with the second term in $p(\xs)$, which, after reordering in powers of
$w$, reads
\eqn\affanpolii{
0\ =\ \sum_{i=1}^{k+1} \tilde{f}_i(y,x,z) w^i = \tilde{f}_{k+1}(y,x,z)\
\big(\, w^{k+1}+\sum_{i=1}^{k}\ f_i(y,x,z)\ w^i \big) \ ,
}
where for $N$ odd
\eqn\effi{\eqalign{
\tilde{f}_i(y,x,z) &= a_{1,i} z^{N+1} + a_{2,i} z^{N-1}x + \dots +
a_{N+1,i} x^{(N+1)/2} \cr
f_i(y,x,z) &= \tilde{f}_i(y,x,z)\ \tilde{f}^{-1}_{k+1}(y,x,z),}}
and similarly for $N$ even.

The functions $f_i$ are rational functions on the torus $E$,
with poles at the zeros of $f_{k+1}$. This is in agreement with the
results in \witex, where the zeros of $f_{k+1}$ are interpreted as
the positions of five-branes on the torus $E$.

\subsec{Affine $D_N$ from Elliptic Fibrations over the Plane}
The second infinite series arises from a base geometry
of 2-spheres intersecting as determined by the affine
Dynkin diagram of $D_n$. The local singularity which is blown up
on the type IIA side is given by
\eqn\adn{
y^2=x^2(x+ct)+t^{N-1},
}
A description of the type IIA geometry based on the standard
representation of the torus as a sextic in $\WP_{1,2,3}$
starts from a toric polyhedron $\Delta$ spanned by the
vertices for the elliptic curve
\eqn\tdn{\ninepoint\eqalign{
E:\rho_1=(0,0,-1),\ \rho_2=(0,-1,0),\ \rho_3=(0,2,3),\ \nu_{a_0}=(0,0,0),}
}
together with $N+1$ vertices describing the blow up
spheres of the singularity \adn:
\def\h{{1 \over 2}\;}
$${\ninepoint\eqalign{
\nu^b_1=(1,1,1),\ &\nu^b_2=(1,2,3),
\nu^a_i=(2,3-i,4-i),\ i=1,\dots,N-3\cr
N \ \ {\rm even}&:\cases{
\nu^c_1=(1,2-n ,2-n ),\cr
\nu^c_2=(1,1-n ,2-n ),\cr}
\cr
N \ \ {\rm odd}&:\cases{
\nu^c_1=(1,1-n ,1-n  ),\cr
\nu^c_2=(1,1-n ,2-n ),}
}}
$$
where $n=\h(N-2)$ for $N$ even and $n=\h(N-3)$ for $N$ odd.
We use again the definition of the moduli \algc\ to
describe the charge vectors $\lm{a}{}$, needed to determine
the mirror geometry:
\def\xg{x^{(g)}}
\eqn\adnrel{{\ninepoint\eqalign{
\zg_i&=\ov{a_ia_{i+2}}{a_{i+1}^2},\ i=1,\dots,N-5,\cr
\yg_1&=\ov{a_2b_1b_2}{a_1^2a_0},\quad
\yg_2=\ov{a_{N-4}c_1c_2}{a_{N-3}^2a_0},\cr
\xg_1&=\ov{a_1\rho_1}{b_1^2},\quad
\xg_2=\ov{a_1\rho_3}{b_2^2},\cr
&\hskip-0.95cm
\cases{\xg_3=\ov{a_{N-3}\rho_1}{c_1^2},\quad \xg_4=\ov{a_{N-3}\rho_2^2\rho_1}
{c_2^2a_0^2}& $N$ even,\cr
\xg_3=\ov{a_{N-3}\rho_2}{c_2^2},\quad \xg_4=\ov{a_{N-3}\rho_2\rho_1^2}
{c_1^2a_0^2}& $N$ odd.}}}}
The elliptic class is given by $(\prod \zg_i)^2(\prod \yg_i)^2(\prod \xg_i)$
with the powers reflecting the Dynkin numbers of the affine root.
The mirror geometry, obtained by solving \lmm, results in the
hypersurface constraint
\eqn\adnpol{
 D_N: \qquad p(\xs)=p_0+vp_1+v^2p_2,}
with 
$$\eqalign{
&p_0= (y^2+x^3+z^6+a_0xyz\big),\cr
&p_1=
(b_1z^Ny+b_2z^{N+3}+c_1z^{4-\ep}yx^{(N+\ep)/2-2}+c_2z^{3+\ep}x^{(N-\ep)/2}),\cr
&p_2=(a_1z^{2N}+a_2z^{2N-2}x+a_3z^{2N-4}x^2+\dots+
a_{N-3}z^8x^{N-4}).}
$$
where $\ep=0\ (1)$ for $N$ even (odd).
The hypersurface constraints $p(\xs)$ are invariant under the
$\ICs$ action $(y,x,z,v)\to(\la^3 y,\la^2 x,\la z,\la^{3-N} v)$
and describe two-complex dimensional Calabi--Yau hypersurfaces,
after appropriate multiplication with powers of $v$.

To construct the 4d superconformal field theory we use this
geometry as the base geometry and then fiber, on the type IIA side,
$A_{n'}$ chains over each 2-sphere in the base, as dictated by the
Dynkin numbers of affine $D_n$, times an overall integer $k$.

\vskip 0.5cm
{\baselineskip=12pt \sl
\goodbreak\midinsert
\centerline{\epsfxsize 2.5truein\epsfbox{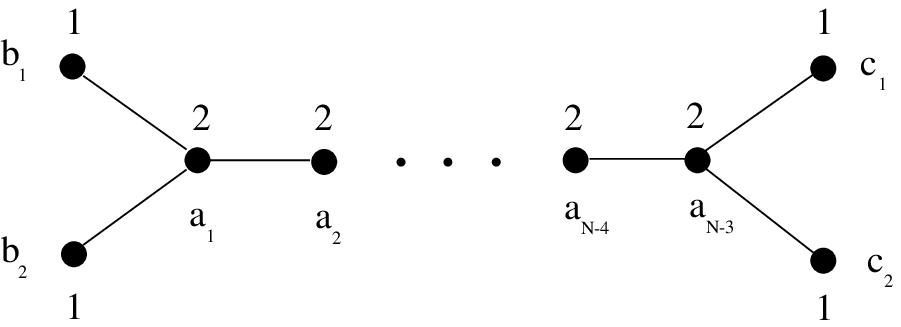}}
\leftskip 1pc\rightskip 1pc\vskip0.5cm
\noindent{\ninepoint  \baselineskip=8pt {\bf Fig. 14:}
Toric diagram for $D_N$ : Dynkin numbers and conventions.
}\endinsert}\ni

As before, we describe these $A_{n'}$ chains by adding  new vertices
to the polyhedron $\Delta$. They are
$$
{\ninepoint
\vbox{\offinterlineskip\tabskip=0pt\halign{\strut
$#$\qquad\hfil &$#$\qquad\hfil &$#$\qquad\hfil\hskip1.5cm &$#$~\hfil \cr
\nu^a_{i,j}=(*,*,*,j),&i=1,\dots,N-3,
&j=0,\dots,2k+1,
&\nu^a_{i,0}\equiv \nu^a_i.\cr
\nu^b_{i,j}=(*,*,*,j),&i=1,2,
&j=0,\dots,k+1,
&\nu^b_{i,0}\equiv \nu^b_i.\cr
\nu^c_{i,j}=(*,*,*,j),&i=1,2,
&j=0,\dots,k+1,
&\nu^c_{i,0}\equiv \nu^c_i.\cr
}}}
$$
and we denote again the moduli multiplying the monomials
of the mirror geometry corresponding to these vertices by
$a_{i,j}$, $b_{i,j}$ and $c_{i,j}$, respectively.

To determine the dimension of $x$, note that we had $N+5$
vertices from the base and we add now $(N-1)(2k+1)+2$ further ones.
Moreover we had $N+1$ coupling parameters measuring volumes
in the base geometry. A 3-fold geometry requires $2(N-1)k+N$
further relations. These are the $2(N-1)k$ Coulomb fields $\zc$
and $N$ mass parameters $\zm$:
\eqn\adnrelii{{\ninepoint
\vbox{\offinterlineskip\tabskip=0pt\halign{\strut
$#$\qquad\hfil &$#$\qquad\hfil &$#$\qquad\hfil\hskip1.5cm &$#$~\hfil \cr
\zc_{a,i,j}=\ov{a_{i,j}a_{i,j+2}}{a_{i,j+1}^2},
&i=1,\dots,N-3,
&j=0,\dots,2k-1\cr
\zc_{b,i,j}=\ov{b_{i,j}b_{i,j+2}}{b_{i,j+1}^2},
&i=1,2,
&j=0,\dots,k-1\cr
\zc_{c,i,j}=\ov{c_{i,j}c_{i,j+2}}{c_{i,j+1}^2},
&i=1,2,
&j=0,\dots,k-1\cr
\zm_{a,i}=\ov{a_{i,0}a_{i+1,1}}{a_{i,1}a_{i+1,0}}
&i=1,\dots,N-4,
&\cr
\zm_{b,i}=\ov{a_{1,0}b_{i,1}}{a_{1,1}b_{i,0}},
&i=1,2,&\cr
\zm_{c,i}=\ov{a_{N-3,0}c_{i,1}}{a_{N-3,1}c_{i,0}},
&i=1,2,&\cr
}}}}
The mirror geometry describing the geometry of the 4d
$N=2$ field theory is then given in terms of the hypersurface
\adnpol\ with 
\eqn\adnpolii{\eqalign{ 
&D_N\ : \ SU(k+1)^4 \times SU(2k+1)^{N-3}\cr
&p_0= (y^2+x^3+z^6+a_0xyz),\cr
&p_1=(b_1z^NyP^{b_1}_{k+1}(w)+b_2z^{N+3}P^{b_2}_{k+1}(w)+\cr
&\hskip 1cm c_1z^{4-\ep}yx^{(N+\ep)/2-2}P^{c_1}_{k+1}(w)+
c_2z^{3+\ep}x^{(N-\ep)/2}P^{c_2}_{k+1}(w)),\cr
&p_2=(a_1z^{2N}P^{a_1}_{2k+1}(w)+a_2z^{2N-2}xP^{a_1}_{2k+1}(w)
+\dots+a_{N-3}z^8x^{N-4}P^{a_{N-3}}_{2k+1}(w)),\cr
}}
where again $\ep=0\ (1)$ for $N$ even (odd) and
the polynomials in the fiber variable are defined as
$$
P^{b_1}_{k+1}(w)=\sum_{j=0}^{k+1}b_{1,j}w^j,$$
and similarly for the other terms.

Eqs. \adnrel\ and \adnrelii\ define the charge vectors $\lm{a}{}$.
The exact solution is then given in terms of the Picard-Fuchs system
\gkz.

\subsec{Moduli space of $G$ bundles on elliptic curves $E$}
We will describe now the relation of the mirror geometry
of elliptic ADE singularities to the geometric representation
of moduli spaces of flat ADE bundles over an elliptic curve.
Physically, these moduli space of $G$ bundles over elliptic curve $E$
is interesting in the light
of the duality between heterotic string on elliptically fibered manifolds
and F-theory. This has been studied in \fmw\ and
\ref\berf{M. Bershadsky, A. Johansen, T. Pantev and V. Sadov,
{\it On four-dimensional compactifications of F theory},
hep-th/9701165}.

Let us collect the equations defining the two complex dimensional mirror
geometry of the base manifold, adding the $E_n$ cases, which can be obtained
in a similar way from fibering the elliptic singularities over the plane:
\eqn\beqs{{\ninepoint
\eqalign{
{\bf A_{N-1}}&:
(y^2+x^3+z^6+\mu yxz)+v(a_1z^N +a_2z^{N-2}x +a_3z^{N-3}y +\dots+
{a_Nx^{N/2}\brace
a_Nyx^{\ov{N-3}{2}}})
,\cr
{\bf D_N}&: (y^2+x^3+z^6+\mu xyz)\cr&+
v(b_1z^Ny+b_2z^{N+3}+c_1z^{4-\ep}yx^{(N+\ep)/2-2}
+c_2z^{3+\ep}x^{(N-\ep)/2})\cr&+
 v^2(a_1z^{2N}+a_2z^{2N-2}x+a_3z^{2N-4}x^2+\dots+a_{N-3}z^8x^{N-4})
,\cr
{\bf E_6}&:(y^2z+x^3+z^3{\zp}^6+\mu xy{\zp}z)+
(a_2v^2x{\zp}^4+a_1vx^2{\zp}^2)\cr&+(b_2v^2y{\zp}^3+b_1vy^2)+
(c_2v^2z{\zp}^6+c_1vz^2{\zp}^6)+d_3 v^3{\zp}^6
,\cr
{\bf E_7}&: (y^2+x^3z +z ^4{\zp}^6+\mu xy{\zp}z)  +
(a_1vx^3+a_2v^2x^2{\zp}^2+a_3v^3{\zp}^4x)\cr&+
(b_3v^3{\zp}^6z +b_2v^2{\zp}^6z ^2+b_1v{\zp}^6z ^3)+
c_2v^2{\zp}^3y +d_4 {\zp}^6v^4
,\cr
{\bf E_8}&:(y^2+x^3+z^6+\mu xyz)\cr&+
(a_6v^6+a_5v^5z+a_4v^4z^2+a_3v^3z^3+a_2v^2z^4+a_1vz^5)\cr&+
(b_3v^3y+b_2v^2x^2+b_4v^4x).
}}}
We assert that the complex deformations of the above two dimensional
surfaces give a geometrical representation of the moduli space of
flat $G$ bundles over an elliptic curve $E$, where $G$ is one of 
the above ADE groups. To this end
note that the divisor $v=0$ projects onto the degree six elliptic curve
$E:\WP^2_{1,2,3}$
with modulus $\mu$. Moreover
the scaling $v\to \lambda v$ induces a projective action on
the moduli parameterizing the complex structure, such that they
become coordinates on the weighted projective spaces:
$${\ninepoint\eqalign{
A_{N-1}:(a_1,\dots,a_N)&\in \IP^{N-1},\cr
D_{N}:(b_1,b_2,c_1,c_2,a_1,\dots,a_{N-3})&\in \WP^N_{1,1,1,1,2,\dots,2},\cr
E_{6}:(a_1,b_1,c_1,a_2,b_2,c_2,d_3)&\in \WP^6_{1,1,1,2,2,2,3},\cr
E_{7}:(a_1,b_1,a_2,b_2,c_2,a_3,b_3,d_4)&\in \WP^7_{1,1,2,2,2,3,3,4},\cr
E_{8}:(a_1,a_2,b_2,a_3,b_3,a_4,b_4,a_5,a_6)&\in \WP^8_{1,2,2,3,3,4,4,5,6}.
}}$$
These are precisely the moduli spaces predicted by Looijenga's analysis
of the moduli space of flat $G$ bundles on $E$ \loo.

Let us describe the geometry of these spaces in more detail. In the
$A_N$ case, $v$ appears only linearly, and integrating it out results
in setting the $v^0$ and $v^1$ pieces to zero separately. The result
is a zero dimensional geometry of $N+1$ points on the elliptic curve $E$.
This is the spectral cover description presented in \fmw \berf. For the $E_n$
cases, setting the extra variable $\zp$ to one in virtue of the present
$\ICs$ symmetries, the equations \beqs\ represent del Pezzo
surfaces\foot{The number in the last bracket
denotes the degree of the polynomial.}
$E_8: \WP^3_{1,1,2,3}[6],\ E_7: \WP^3_{1,1,1,2}[4],\ E_6: \WP^3_{1,1,1,1}[3]$,
respectively.  Again this agrees with the representation of $E_n$ bundles
in terms of complex deformations of del Pezzo surfaces in ref. \fmw.

This identifies the apparently disconnected descriptions of the geometrical
objects appearing in the analysis of \fmw\ as the mirror
of the {\it physical type IIA
compactification
geometry}. Note that the reasoning given in section 3 explains
the connection we have found between the mirror of this type IIA
geometry and the moduli of flat ADE bundles on an elliptic curve.
Moreover here we have obtained also the geometrical description for the $D_N$
case for which no representation in terms of complex deformations
was known. As mentioned, the non-simply laced cases can be obtained in
the same way \pape.

\subsec{S-duality Groups}

As discussed in the previous sections, when
we consider the chain of $SU$ groups
arranged according to the affine ADE Dynkin diagrams
whose ranks are proportional to the Dynkin indices and
 where for each link we associate bi-fundamental matter,
we obtain an $N=2$ theory in four dimensions
with vanishing $\beta$-function.  Similarly if we consider
the configuration of $SU$ groups according to ordinary
ADE diagram with extra matter fields as described previously, we
obtain again superconformal theories.
Thus the coupling constants
of these gauge groups do not run
and it is natural
to ask what is the space of inequivalent coupling
constants.  This moduli space is the naive
classical moduli space of coupling constants
modulo the S-duality group.

{}From our construction of the type IIA geometry
it is clear that the space of couplings
is the same as the moduli controlling the (affine) ADE
geometry of the base.
Moreover the moduli space of the affine base geometry describes the
blow up of elliptic ADE singularities and is
equivalent to the moduli $\cx M_G$ of flat $G\subset ADE$ bundles on a 2-torus $E$,
as we discussed in section 3 and the present section.
{\it The S-duality group is then simply the fundamental group
of $\cx M_G$}.
In the case of ordinary ADE diagrams, the
S-duality group can be obtained by degenerating the moduli
of the elliptic curve in the corresponding affine case and
the S-duality group is the subgroup of the
affine one, which corresponds to moduli of flat ADE
connection on the degenerate elliptic curve.

The case corresponding to the $A$ diagrams was already
considered in \witex\foot{See also 
\ref\argn{P. Argyres, {\it S-Duality and Global Symmetries
in $N=2$ Supersymmetric Field Theory}, hep-th/9706095}.}, where
the moduli space was shown to be the moduli of $n$ points
on an elliptic curve or its degeneration
depending on whether we are dealing
with the affine or ordinary $A$ case.  This is in agreement with our
result when one notices that flat bundles of $A$-type on a torus
are equivalent to the choice of
$n$ points on the dual torus or its degeneration.

It is quite suggestive
that a gauge group associated with the base
arises in describing this moduli space.  This begs
for a more direct physical interpretation, and is
related to the strong coupling phenomena
associated to shrinking the base, as we will discuss
now.

\newsec{Strong Coupling Fixed Points and New Dualities}
We have seen that in our construction the base and fiber
play a similar role.   For example consider the curve
we have for the linear chain of $SU(n)^m$ gauge theories
along a linear chain with $n$ additional fundamentals
at each end of the chain.  The corresponding threefold
for this case is given by
\eqn\duagp{
F(z,w)=\sum_{i=0}^{m+1}\sum_{j=0}^{n} \;
a_{i,j}\;  z^i\;  w^j = uv,}
where $z$ corresponds to the base degree of freedom and $w$ to the fiber,
as before. Clearly this geometry is invariant
under the exchange of $n\leftrightarrow
m+1$ and $z\leftrightarrow w$, which would correspond to the geometry
associated with $SU(m+1)^{n-1}$ along the linear chain with
$m+1$ extra fundamentals at each end.  In particular
in either case the relevant fivebrane lives on the same genus $m\; (n-1)$
Riemann surface. This suggests a ``duality'' of the form
\eqn\duag{SU(n)^m \leftrightarrow SU(m+1)^{n-1}}
with the matter described above, where the Coulomb
parameters, the couplings and the masses get exchanged in a
non-trivial fashion.

\vskip 0.5cm
{\baselineskip 12pt\sl
\goodbreak\midinsert
\centerline{\epsfxsize 1.2truein\epsfbox{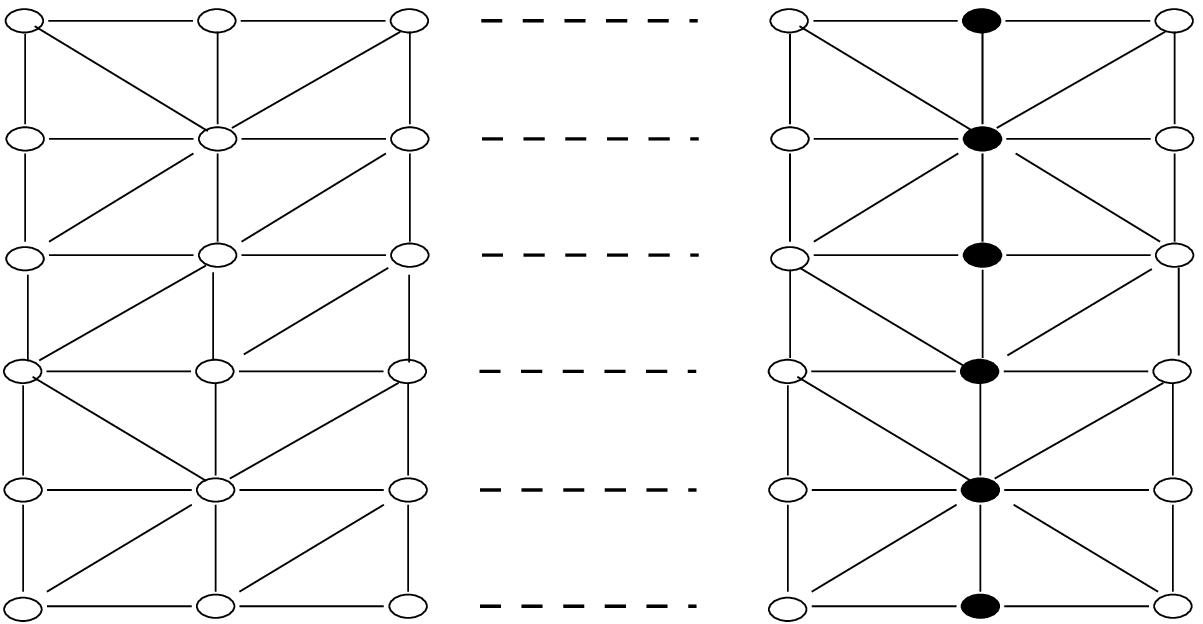}}\leftskip 1pc
\rightskip 1pc\vskip0.3cm
\noindent{\ninepoint  \baselineskip=8pt {\bf Fig. 15:}
Toric diagram for the field theories \duag\ connected by
the strong coupling fixed point.
}\endinsert}\ni

Before we analyze the relation in \duag\ in more detail,
let us use it to get some more insight about the S duality group
of each one of these theories. The S-duality group describes
the equivalences of the moduli space in the high energy
behavior, or equivalently, the behavior for small values
of the moduli. Neglecting the terms of
lower degree in $w$ in \duagp\ (for the $SU(n)^m$ theory),
$F(z,w)/w^n$ is of the form
$$
\sum_{i=0}^{m+1} a_{i,n}z^{i}.
$$
The moduli space is described by the $m$ Coulomb
moduli\foot{More precisely there are finite shifts proportional
to the mass parameters which are straightforwardly to determine
from eqs. \aco.}
$\zc$ of the $SU(m+1)$ theory  corresponding to the base
geometry. In particular, the effective duality
group acting on the Coulomb parameters of $SU(m+1)$
generates the S-duality transformations of the $SU(n)$ theory.

As an explicit example consider the $m=1$ case. The coupling
space describes the roots of an quadratic equation
which, written in terms of the $SU(2)$ modulus
$\zc=a_{2,n}a_{0,n}a_{1,n}^{-2}$
reads $F/w^n=z'^2+(4\zc-1)$.
The singular points are at the values $\zc=0,\ov{1}{4},\infty$.
The singularity $\zc=0$, the large base limit,
corresponds to weak coupling limit whereas
$\zc=\infty$ is a $\IZ_2$ orbifold point, as discussed in
\witex\ in the M-theory picture. The five brane
description breaks down at the point $z=\ov{1}{4}$
corresponding to a collapse of two five branes.
In the type IIA geometry, this singularity describes simply
the zero volume point of the blow up sphere of the $A_1$
base geometry.

In the general $SU(m+1)$ case there are $m$ singular points
of this type
in the generic hyperplane of the moduli space. The S-duality
group is generated by loops around these singular points together
with the weak coupling monodromies generating the translations
in the special coordinate $t\sim \ov{1}{2\pi i}\ln \zc$.
Analogous statements apply to the coupling space of the
$D$ and $E$ cases, as is clear from the above discussion
and also from the large $w$ limit of the polynomials
given in the previous sections.

Let us determine now more carefully the precise
relation between the theories appearing in \duag.
The local geometry of the type IIB side and its periods
describe part of the exact moduli space of the $N=2$
string theory, which is of special K\"ahler type. To reduce
to the field theory we have on the one hand to decouple
gravity. Moreover also the bare parameters of the field
theory, the coupling constants and the mass parameters,
are scalar fields which sit in full vector multiplets.
Clearly we have to freeze out the fluctuations of these
fields, if we are interested in the pure field theory
answer described by an action including renormalizable
couplings only.

In the asymptotic free case, the $M_{pl}\to \infty$ limit
requires to adjust the coupling constants to zero at the
Planck scale \kklmv, in order to keep the field theory
scale $\Lambda$ at a
fixed, finite value. Moreover, the vector multiplets corresponding
to the bare parameters freeze out due to the behavior of
their kinetic terms, whereas the vector fields of the
gauge theory remain dynamical, if the moduli are tuned to
to the neighborhood of the enhanced gauge symmetry point
in moduli space.

On the contrary, in the case with vanishing $\be$ function, the
couplings do not run above the natural scale of the field theory
set by the vev's and the volume of the base remains
unfixed in the $M_{pl}\to\infty$ limit. We have still to adjust the
field theory moduli to be close to the enhanced symmetry point.
However in this case it is possible to treat the fiber
geometry in the same way as the base
and we obtain a {\it new} low energy theory containing
the coupling constants as additional dynamical fields. In particular
note that although the curve \duagp\ is as expected in gauge theory,
the differential $\Omega$ is now symmetric in the
base variable $z$ and the fiber variable $w$ and
agrees with the field theory answer only after a shift the
variable $w\to \cx O(\ep^{-1})+w$:
$$\eqalign{
\Omega &= \ln(w)d\ln(z)= - d\ln(w)\ \ln(z)\cr
&=-\ln\ep\ov{dz}{z}+\ep w\ov{dz}{z}+\sum (-)^{k-1}\ep^k\ov{w^k}{k}\ov{dz}{z}}
$$
In the asymptotic free case, the first term corresponding to
the period associated with the base volume is related to the
weak gravity limit by $e^{-S}=(\Lambda^2\al')^l=\ep^l$,
where $S$ is the dilaton determining the string coupling,
$\al' \sim M_{string}^{-2}$ and
$l$ depends on the theory we consider. The 1-form proportional
to $\epsilon$ in the above expression is what the gauge system sees.
 In the conformal case,
the exponent $l$ is zero and
the scale $\ep$ is a new scale governing the dynamics
of the additional fields from the base.
It is clear that we can now switch the roles of the base and the
fiber in the above limit. In fact we can continuously interpolate
between the theories \duag.

Of course these new theories symmetric in
the base and fiber geometry are very interesting to study further.
In order to gain insight into these cases, let us consider the
closely related, but simpler case considered in
\ref\dkv{M. R. Douglas, S. Katz and C.  Vafa,
{\it Small instantons, Del Pezzo surfaces and
type I' theory}, hep-th/9609071}\ref\ms{D. R. Morrison and N. Seiberg, \nup 483
(1997) 229}\kkv:  If we consider ${\bf P}^1\times {\bf P}^1$ in
a Calabi-Yau,  and view one of the ${\bf P}^1$'s as the
fiber, we obtain a theory with pure $SU(2)$ gauge symmetry
where the volume of the base ${\bf P}^1$ is related to the
gauge coupling of the $SU(2)$, $V\sim 1/g^2$.   However
if we exchange the role of base and fiber we get the base
${\bf P}^1$ giving rise to $SU(2)$ and the volume of the
fiber ${\bf P}^1$ is related to its coupling.
So in a sense we have an $SU(2)\times SU(2)$ here, where the
Coulomb space of either $SU(2)$ is identified with the
coupling constant of the other $SU(2)$.  This suggests couplings roughly
of the form
$$\tr F_1^2 f(\phi_2) +(1\leftrightarrow 2),$$
where $\phi_i$ denote the scalar adjoints of the two $SU(2)$'s and
where $f$ vanishes as $\phi_2\rightarrow 0$.
This theory is clearly not renormalizable, but the critical point
corresponding to a superconformal theory where roughly speaking
we are at the origin of the Coulomb branch for both $SU(2)$'s is
known to exist \dkv\ms.

The situation we are considering above is roughly of the same
type, where now the Coulomb parameters of the base $A_m$
play the role of the couplings of the various $SU$ gauge
groups in the fiber.  But now the classical Weyl group of $A_m$
exchanges the various $SU$ gauge groups with each other
thus the coupling of these two systems makes sense only
if we consider the part of the S-duality group which
exchanges the $SU$'s.  So in some sense
it is like gauging the quantum symmetries of the theory.
Clearly this is a very interesting area to study further.

Note also that the affine ADE bases that
we have considered, if we allow shrinkings of the base, would correspond
to new critical theories, and in fact correspond to compactification
of M-theory on the same manifold times a circle, or F-theory
on the elliptic version of the same manifold times a 2-torus
\ref\moris{K. Intriligator, D. R. Morrison and N. Seiberg,
{\it Five-dimensional supersymmetric gauge theories and
degenerations of Calabi-Yau spaces}, hep-th/9702198}%
\ref\bva{M. Bershadsky and
C. Vafa,
{\it Global anomalies and geometric engineering of critical
theories in six-dimensions}, hep-th/9703167}%
\ref\inbl{J. D. Blum and K.
Intriligator,
{\it Consistency conditions for branes at orbifold
                  singularities}, hep-th/9705030;
{\it New phases of string theory and 6-D RG fixed points via
                  branes at orbifold singularities},
hep-th/9705044}%
\ref\mas{
P. Aspinwall and D. R. Morrison,
{\it Point - like instantons on K3 orbifolds},
hep-th/9705104}%
\ref\ln{A. Lawrence and N. Nekrasov,
{\it Instanton sums and five-dimensional gauge theories},
hep-th/9706025}.
Thus our results give answer to the Coulomb branch of
such 6  dimensional critical theories compactified on $T^2$.

\newsec{Applications to $d=3$, $N=4$ QFT's}
Some of the results we have obtained  have been used in
\hov\ to derive dual pair of field theories with $N=4$ in $d=3$.
This is done by considering a further compactification
of our mirror IIB model and considering the Higgs branch
and using T-duality on the extra circle
and converting it to a IIA model Coulomb branch,
compactified on a circle, and reading off the gauge
theory and matter content.  Even though in this paper
we have mostly concentrated on theories which
are asymptotically free in $d=4$ (ignoring
the $U(1)$ factors), we already mentioned how
one would construct the corresponding geometry even for
non-asymptotically free theories.
For applications in three dimensional QFT's
the cases which are not asymptotically free in $d=4$
(but which automatically are asymptotically free in $d=3$)
are more relevant as
those are the cases that
in cases which is
completely Higgsable and which can have a dual gauge theory
with Higgs and Coulomb branches exchanged \inse
\ref\oogho{
J. de Boer, K. Hori, H. Ooguri and Y. Oz,
{\it Mirror symmetry in three-dimensional gauge theories,
                  quivers and D-branes}, hep-th/9611063;
{\it Mirror symmetry in three-dimensional theories, SL(2,Z) and
                  D-brane moduli spaces}, hep-th/9612131}%
\ref\hanw{A. Hanany and E. Witten,
{\it Type IIB superstrings, BPS monopoles, and three-dimensional
                  gauge dynamics}, hep-th/9611230}.

It is clear from the approach in \hov\ that a dual
system for any $N=4$, $d=3$ gauge system will exist
if it comes from geometric engineering, however
the dual may not be a gauge system.  This in particular
was shown to be the case for the dual of $\prod U( s_i)$ groups
associated to an affine $E_{8,7,6}$ Dynkin diagram, where
$s_i$ are the Dynkin indices.  The dual in this case was
found to be the toroidal compactification of one $E_{8,7,6}$ small
instanton in accordance with the conjecture \inse.
The method used there was to start from the Coulomb branch
of the compactified exceptional strings in the type IIB setup
which was known
\ref\relre{
J. A. Minahan and D. Nemeschansky, \nup 482 (1996) 142;
 \nup 489 (1997) 24;\br
O. Ganor, \nup 488 (1997) 223;\br
O. Ganor, D. R. Morrison and N. Seiberg, \nup 487 (1996) 93;\br
W. Lerche, P. Mayr and N. P. Warner,
{\it Noncritical strings, Del Pezzo singularities and Seiberg-
                  Witten curves, hep-th/9612085}
}
and then considering
the Higgs branch of it and reading it as type IIA theory.
However we can study the same problem in a completely different
way, using the results of the present paper. Namely, we can
start from the type IIB realization of the Coulomb branch
of the $SU$ groups associated with the affine $E$ quivers
and see if that can be viewed as the Higgs branch of the
small $E_n$ instantons, which is known
\ref\wit{E. Witten, \nup 471 (1996) 195}%
\ref\mvii{D. R. Morrison and C. Vafa, \nup 476 (1996) 437}
for $E_8$ theory, but the generalization to the other cases
goes through without any complications.

Note that here it is crucial that we are actually dealing
with $\prod U(s_i)$ rather than $\prod SU(s_i)$.  As noted before
the extra $U(1)$'s do not affect the Coulomb branch we studied
in four dimensions and in fact the vev of the scalars in the $U(1)$
correspond to the mass parameters of the bi-fundamental matter.
However they are crucial in the three dimensional story as the
$U(1)$'s are asymptotically free in $d=3$.  Let us recall
the local type IIB geometry which gives the
Coulomb branch of this theory  for the more general case of $\prod U(ks_i)$
\affenallpol :
$$\eqalign{E_8:\ p(\xs) =&y^2+x^3+z^6+\mu xyz\
+\sum_{i=1}^6z^{6-i}P^{z}_{i\cdot k}(w)\ + \cr
&\sum_{i=1}^2x^{3-i}P^x_{2i\cdot k}(w)
+\sum_{i=1}^1y^{2-i}P^y_{3i\cdot k}(w)}$$
Let us restrict first to $k=1$.  Furthermore, we can
shift $x,y$ so that the equation will involve only the
monomials $y^2,x^3,x,1$, with coefficients a function of
$z,w$.  Once we do this we find that the local geometry can be
written as
$$\eqalign{y^2=&x^3+z^6+a xz^4+x(z^3 f_1(w)+z^2 f_2(w)+zf_3(w)+f_4(w))\cr&
+(z^5g_1(w)+z^4 g_2(w)+...+g_6(w))}$$
where $f_i(w),g_i(w)$ are polynomials of degree $i$ in $w$
(which can be written in terms of the original polynomials).
This is exactly the description of the Higgs branch of one small $E_8$
instanton in F-theory, which when compactified on $T^2$
becomes the description of type IIA on the same geometry,
thus showing that the dual of one $E_8$ instanton compactified
to three dimensions is the $\prod U(s_i)$ along the
affine $E_8$ quiver diagram.  The generalization to
$U(ks_i)$ is straightforward:  All that happens is that one
shifts the degrees of
polynomials $f_i,g_i\rightarrow f_{k\cdot i},g_{k\cdot i}$
in the above equation, and that is what one expects for
$k$ instantons of $E_8$
\ref\mvi{D. R. Morrison and C. Vafa, \nup 473 (1996) 74}\mvii .
This is also in accord
with the conjecture \inse.

Note that our approach will also allow us to find new dual
systems.  For example suppose we are interested in
constructing the dual to $U(s_i)$ along the affine $E_8$
Dynkin diagram and in addition some extra fundamental
matter for each group.  Then the methods of the previous
section can be used to give the geometry for the 3d dual type IIA
Higgs branch, just as in the case considered. Then, however, it will
not be related to any known small instantons theory, but the dual system
will always exist, as was noted in \hov .

We would like to thank N. Elkies, K. Hori, R. Gebert, 
W. Lerche, H. Ooguri and N. Warner for valuable discussions.
C.V. would also like to thank
the hospitality of Institute for Advanced Study.

The research of S.K. was supported in part by NSF grant DMS-9311386
and NSA grant MDA904-96-1-0021.
The research of P.M. was supported by NSF grant
PHY-95-13835. The research of C.V. was supported in part by NSF grant
PHY-92-18167.

\listrefs
\end